\documentclass{article}
\usepackage{authblk}

\usepackage[T1]{fontenc}

\usepackage[utf8]{inputenc}
\usepackage{geometry}
\usepackage{changepage}

\usepackage{comment}

\usepackage{textcomp}
\usepackage{float}
\usepackage{amsmath, amssymb}
\usepackage{braket}
\usepackage[english]{babel}
\usepackage{dirtytalk}
\usepackage[separate-uncertainty=true]{siunitx}
\usepackage{graphicx}
\usepackage[labelfont=bf]{caption}
\usepackage{subcaption}
\usepackage{mathtools}
\usepackage{physics}
\usepackage{amssymb}
\usepackage{amsmath}
\usepackage{mathtools}
\usepackage{xfrac}
\usepackage{array}
\usepackage{relsize}
\usepackage{tabu}
\usepackage{bbold}
\usepackage{tensor}

\usepackage[font=small]{caption}
\usepackage{amsfonts}
\usepackage{booktabs}
\usepackage{siunitx}
\usepackage{appendix}
\usepackage{mathrsfs}
\usepackage{amsthm}
\usepackage{indentfirst}
\usepackage{soul} 
\setstcolor{orange}
\usepackage{xcolor}
\usepackage{hyperref} 
\usepackage{xcolor} 
\setstcolor{cyan}
\usepackage{booktabs}
\usepackage{makecell}

\usepackage{phfcc}
\phfMakeCommentingCommand[initials={R.P.}]{RP}      % RP
\phfMakeCommentingCommand[initials={M.F.}]{MF}
\phfMakeCommentingCommand[initials={S.C.}]{SC}
\DeclareSIUnit\parsec{pc}
\DeclareSIUnit\century{century}

\numberwithin{equation}{section}
\numberwithin{figure}{section}

\allowdisplaybreaks[1]
\usepackage{systeme}
\usepackage{tensor}

\title{Local observers in stationary axisymmetric dust spacetimes\footnote{We dedicate this work to the memory of Prof. Vittorio Gorini (1940-2025), who has been a constant guide and inspiration for all of us. Vittorio, with his deep enthusiasm, wit and clarity, saw this work since its inception and supported it until its departure. Have a good trip, Vittorio!}}
\author{Matteo Fontana, Sergio Luigi Cacciatori, Roberto Peron}

\author{Matteo Fontana$^{1,2}$\footnote{mfontana11@uninsubria.it}, Sergio Luigi Cacciatori$^{1,2,3}$\footnote{sergio.cacciatori@uninsubria.it},
Roberto Peron$^{4}$\footnote{roberto.peron@inaf.it}\\ 
$^{1}$DiSAT, Universit\`a dell'Insubria, via Valleggio 11, 22100 Como, Italy\\ 
$^{2}$Istituto Nazionale di Fisica Nucleare (INFN), Sezione di Milano, via Celoria 16, 20133 Milano, Italy\\
$^{3}$Como Lake centre for AstroPhysics (CLAP), DiSAT, Università dell’Insubria, via Valleggio 11, 22100 Como, Italy\\
$^{4}$Istituto Nazionale di Astrofisica (INAF), Istituto di Astrofisica e Planetologia Spaziali (IAPS), Via del Fosso del Cavaliere 100, 00133 Roma, Italy\\
}

\date{}

\begin{document}

\maketitle

\begin{abstract}
    In this work, we construct a locally inertial reference system adapted to a geodesic observer in stationary, axisymmetric dust solutions of the Einstein equations employed as effective models of a portion of a galactic disc. To ensure a consistent spatial orientation among different local observers, we also introduce the radially locked reference system, in which one spatial axis is aligned with the radial direction defined by null geodesics passing through the galactic center. Within this framework, we analyze how the dust configuration is described by such observers by computing the frequency shift of photons exchanged between pairs of dust geodesics. Building on this construction, we outline a procedure to reconstruct spectroscopic and astrometric relative velocities with respect to locally inertial observers, providing a coherent foundation for the study of galactic kinematics in a fully general relativistic context.
\end{abstract}

\section{Introduction}
\label{Introduction}

General Relativity (GR) is, to date, the most precise and well-tested theory of gravitation~\cite{Will:2014kxa,2008ARNPS..58..207T}. It has consistently passed all experimental and observational challenges, from Solar System dynamics to the detection of gravitational waves~\cite{LIGOScientific:2016aoc} and the study of black holes at galactic centers~\cite{EventHorizonTelescope:2019dse}, with remarkable accuracy. In regimes of weak gravitational fields and low velocities, GR is expected to reduce to Newtonian gravity, with relativistic corrections that can be systematically incorporated through an expansion in powers of $v/c$, where $v$ is the characteristic velocity of the system and $c$ the speed of light~\cite{PoissonWill2014}. This has justified the longstanding use of Newtonian mechanics to model galactic dynamics, as galaxies typically exhibit low velocities and weak gravitational potentials. However, a critical discrepancy emerges: Newtonian gravity fails to explain the observed dynamics of galaxies. In particular, the rotational velocities of stars and gas in the outer regions of disc galaxies do not decline as expected from the visible mass distribution, but instead remain approximately constant with increasing radius, giving rise to the well-known flat rotation curves~\cite{Rubin:1970zza,Bosma:1981zz,SofueRubin2001}. This non-Keplerian behavior may point either to a limitation in the theory of gravitation being used or to the need for additional matter beyond what is observed~\cite{Cacciatori2024}\footnote{This is a well known pattern in the history of science: think about the \SI{43}{\arcsecond\per\century} residual in the perihelion precession of planet Mercury orbit around the Sun, known since 1859, well before the introduction of GR. Possible explanations to this discrepancy included a further, heretofore unknown, planet and a modification of the Newtonian law of universal gravitation.}. The dominant view has been to uphold Newtonian gravity as the appropriate weak-field, slow-motion limit of GR, while postulating the existence of a non-baryonic form of matter interacting primarily gravitationally with the ordinary Standard Model particles. This hypothetical component, known as dark matter (DM), is invoked to account for the observed flat rotation curves by surrounding galaxies with extended halos possessing suitable density profiles~\cite{Navarro:1996gj}. Beyond galactic rotation curves, DM has been used to explain a broad range of astrophysical and cosmological phenomena that appear to require more gravitational mass than what is directly observed, becoming a central component in the prevailing $\Lambda$ Cold Dark Matter ($\Lambda$CDM) cosmological model. 
Supporting evidence includes virial estimates of galaxies within galaxy clusters~\cite{Zwicky:1933gu,Zwicky:1937zza,lopez}, gravitational lensing patterns~\cite{Massey:2010hh}, thermodynamic emissions of X-rays in galaxy clusters~\cite{Castillo-Morales:2003mdk}, the dynamics of the Bullet Cluster~\cite{Clowe:2003tk,Markevitch:2003at}, measurements of the deceleration parameter~\cite{Planck:2018vyg}, large-scale structure formations~\cite{Springel:2005nw}, and the peaks observed in the Cosmic Microwave Background~\cite{Dvorkin:2022bsc}.
A wide variety of DM candidates have been proposed, ranging from weakly interacting massive particles and axion-like particles to massive compact halo objects~\cite{Bertone:2004pz}. Yet despite decades of theoretical modeling and experimental effort, the precise nature of DM remains elusive. Alongside the challenges posed by dark energy, this quest to comprehend the dark sector invites a reexamination of the $\Lambda$CDM model and its implications for our understanding of the Universe~\cite{CacciatoriE2024}.

Importantly, all current evidence for DM comes from gravitational phenomena\footnote{It is interesting to notice that most of the information which can be gained on gravitational dynamics comes from electromagnetism, i.e. from the light we receive from distant objects. Indeed, one can trace the origin of DM models in the context of galactic dynamics in an attempt to reconcile the knowledge of galactic luminosity with that of mass density distribution (see e.g.~\cite{1977PNAS...74.1767O}).}. This motivates an alternative approach: the mismatch between visible mass and observed dynamics might point not to unseen matter, but to a breakdown of our current theory of gravity on galactic and cosmological scales.
This insight led to the formulation of Modified Newtonian Dynamics (MOND)~\cite{Milgrom:1983ca, Milgrom:1983pn, Milgrom:1983zz}, suggesting that Newton's law of gravitation ceases to hold in regions of very low acceleration (below a certain scale $a_0 \sim 10^{-10}\text{m/s}^2$), like galaxy edges, where gravity would weaken more slowly than predicted by the inverse-square law. This empirical modification can explain flat rotation curves without DM and has had notable success on galaxy scales with a single additional parameter $a_0$~\cite{FamaeyMcGaugh2012}. However, MOND is not a relativistic theory by itself, and while relativistic extensions exist, they face challenges, particularly in matching cluster- and cosmological-scale observations and in providing a consistent description of gravitational lensing and cosmic microwave background data~\cite{FamaeyMcGaugh2012}.

Another approach is provided by modified gravity theories (MOGs) \cite{CANTATA:2021asi}, which seek to extend or revise GR by altering the Einstein–Hilbert action, either through modifications to the geometric sector or by introducing additional fields that mediate the gravitational interaction. These models aim to explain cosmic acceleration and galactic dynamics without invoking DM or dark energy. For instance, some $f(R)$ models can produce galaxy-scale effects that mimic DM by effectively modifying the Newtonian potential in the weak-field, slow-motion limit~\cite{Capozziello:2006dj}. However, these theories often come with extra degrees of freedom and parameters, whose choices can be somewhat arbitrary, and they must be carefully tuned to satisfy the stringent experimental tests of GR in the Solar System~\cite{DeFelice:2010aj}. Furthermore, in their weak-field limit, they still rely on essentially Newtonian concepts.

However, since the currently accepted theory for gravitation, i.e., GR, already departs significantly from Newtonian theory, it is worthwhile to explore the potential of fully general relativistic models of galaxies. Galaxies, unlike the Solar System, are large-scale, rotating mass distributions, and the assumptions underlying the post-Newtonian framework, while highly successful in describing Solar System dynamics, may not readily extend to such regimes. This is due to the fact that GR possesses features absent in the Newtonian theory, which could, in principle, cumulatively affect dynamics over large scales. Notably, the Einstein equations (EE) are nonlinear, and GR encompasses a broader set of dynamical degrees of freedom, encoded in the spacetime metric, compared to the single scalar gravitational potential of Newtonian gravity. An important example is the gravitomagnetic frame-dragging effect arising from off-diagonal components of the metric that can be comparable in magnitude to the Newtonian potential in stationary axisymmetric systems~\cite{Astesiano:2022ghr, Ruggiero:2023geh, Ludwig:2024mgo, Astesiano:2024zzz}. Direct evidence for the actual existence of frame-dragging effects has been accumulating in the past decades, based on the advances in space geodesy techniques~\cite{1995grin.book.....C,2006NewA...11..527C,2019EPJC...79..872C,2020Univ....6..139L}. GR also permits nontrivial global topology, allowing for spacetimes that are locally flat but globally conical, exhibiting angular deficits or excesses far away from the rotation axis \cite{Fontana:2024aru}. Such a topological feature has no analogue in Newtonian gravity, which is formulated on a flat Euclidean space. These considerations suggest that, even in weak-field, slow-motion regimes, general relativistic effects that cannot be treated as small corrections to Newtonian gravitation, might play a significant role. It is therefore worthwhile to seek exact solutions of the EE suited to modeling extended rotating systems, in order to identify genuine relativistic contributions that survive in this regime and might account for the observed phenomena beyond the reach of Newtonian gravitation alone. \\
Finding exact, physically viable solutions of the EE for realistic galaxies is notoriously difficult. In the general case of a rotating, extended mass distribution, the resulting equations are analytically intractable without imposing strong symmetry and matter assumptions. A widely used simplification is to model a galaxy as a stationary axisymmetric solution of the EE sourced by a pressureless perfect fluid (dust). This approximation is expected to hold in a portion of the galactic disc, which exhibits approximate axial symmetry and where the mean time between stellar encounters greatly exceeds the galaxy’s age, rendering collisional effects negligible~\cite{BinneyTremaine2008}. Early attempts to build galaxy models of this sort~\cite{Cooperstock2005,Cooperstock:2006dt,Cooperstock:2006ti,Cooperstock:2007sc} sparked controversy by claiming that GR effects might explain flat rotation curves without DM. However, those solutions suffered from unphysical features, such as violation of the weak energy condition on the galactic plane. A step forward came with the Balasin-Grumiller (BG) model~\cite{BG}, which describes a rigidly rotating, thin dust disc. This framework was subsequently extended to allow for differential rotation of the dust, representing more faithfully the dynamics of a galaxy in which stars at different distances from the rotation axis move with different angular velocities~\cite{Astesiano:2021ren}. The resulting class of solutions, known as $(\eta,H)$ models, is characterized by an arbitrary function $\eta$ and a related negative function $H(\eta)$. It encompasses the full set of stationary axisymmetric dust solutions to the EE~\cite{Geroch:1970nt,Geroch:1972yt,Winicour,Stephani:2003tm}, including BG as a limiting case. These generalizations have been inspired by recent studies which, beyond the DM distribution in Newtonian approximation and MOND, tested the BG velocity profile relative to zero angular momentum observers (ZAMOs) with Gaia DR2 and DR3 data for the Milky Way, spanning radii between 5 and 19 kpc, and showed that it reproduces the observations without invoking DM, yielding results statistically indistinguishable from state-of-the-art halo models~\cite{Crosta:2020,Beordo:2024}. These results suggest looking for more general solutions yielding explicit expressions of the velocity of dust with respect to a properly defined class of observers, which could be interpreted as describing galaxy rotation curves.

It is important here to stress that two different types of problems are to be faced. In the \emph{dynamic} one, some type of law of motion for the galactic components (e.g., stars) is to be found from the EE. In the \emph{kinematic} one, the focus is shifted to the observation of this motion from a specific point of view, which is in practice inside the Solar System. As we are going to see, in a general relativistic setting, these two problems are inextricably connected due to the dynamic nature of spacetime.

In standard astronomical practice, observations are referenced to the Barycentric Celestial Reference System (BCRS)~\cite{Soffel_2003,2005USNOC.179.....K,1990CeMDA..48..167H}. This system is centered at the Solar System barycenter, with spatial axes fixed relative to distant, kinematically non-rotating sources. Its underlying metric relies on a post-Newtonian approximation that treats the Solar System as an isolated entity. However, applying this framework to exact general relativistic galaxy models presents two fundamental challenges. First, the metric of a stationary axisymmetric dust configuration introduces relativistic effects that cannot be captured by a standard post-Newtonian expansion, rendering the `isolated Solar System' assumption inconsistent. While one might attempt to address this point by constructing a reference system in the full theory adapted to static observers with asymptotically fixed axes \cite{Costa2023}, a second, critical issue remains. Since GR is inherently a local theory, defining asymptotically fixed spatial directions implicitly requires the solution to extend to infinity. 
This extrapolation is physically unjustified for galaxy models, which are intended as effective descriptions valid only within a restricted spacetime domain. 
Moreover, it is worth noting that the need for asymptotic structure is not unique to galactic models and also underlies the relativistic description of the Solar System. The crucial difference, however, is the presence of a clear scale separation in the latter case: the Sun dominates the gravitational field and the weak-field regime allows for a controlled post-Newtonian expansion around an effectively isolated system. Galactic models, by contrast, describe collective self-gravitating matter distributions without a comparable hierarchy or controlled asymptotic regime. In this setting, the assumption of asymptotically fixed spatial directions is much less physically justified. \\
In this work we investigate the interpretation of kinematic observations within stationary axisymmetric dust solutions of the EE, employed as effective models for a portion of a galactic disc describing averaged motions of its constituents. The analysis is carried out in a fully local framework, without reference to distant non-rotating objects. We construct a dynamically non-rotating reference system associated with locally inertial observers, which can be operationally realized by a triad of mutually orthogonal gyroscopes undergoing Fermi–Walker transport along the observer’s worldline. This construction should be understood as an effective, averaged description of motion within the galactic disc, since it neglects the proper motions of individual stars and other substructures. Nevertheless, it already highlights the intrinsic difficulty of interpreting kinematic observations in GR and underscores the essential differences with the Newtonian framework commonly adopted in galactic dynamics.

In the Newtonian framework, time is absolute and space is treated as a fixed Euclidean background in which dynamics unfolds. Within this setting, inertial coordinate systems cover the entire spacetime and are related by Galilean transformations, which leave the physical description invariant. They agree on the notions of simultaneity and spatial distance, and free particles follow straight-line trajectories indefinitely. This universal inertial structure underlies the classical treatment of dynamics and enables a straightforward interpretation of astronomical measurements.

In GR, the situation changes fundamentally. The curvature of spacetime prevents the existence of globally inertial reference frames. Inertial frames can only be defined locally, within regions small enough that tidal effects are negligible\footnote{This is the essence of the Equivalence Principle as Einstein conceived it in the formulation of GR.}. Beyond these neighborhoods, the influence of spacetime curvature becomes significant, and inertial motion cannot be globally defined. As a result, the interpretation of motion becomes inherently local, and comparisons between different spacetime events must be treated with care~\cite{MisnerThorneWheeler1973,Wald:1984,1999grga.book.....L}.

A central challenge is then determining how to consistently compare measurements performed by distinct local observers, each operating within a limited region where their reference frame is valid. This issue is particularly relevant in astronomy, where observational quantities, such as velocities and angles, must be referred to well-defined spatial directions. A possible approach involves exploiting the propagation of light to establish a common directional standard between different local frames. This leads to the concept of a radially locked reference system: one constructed by rotating the spatial triad of a locally inertial observer so that one axis aligns with the direction of radial null rays, as perceived within the local frame. Such a construction provides a physically meaningful way to match local observations across curved spacetimes.

Finally, it is important to recognize that galactic rotation curves, which express the radial velocity as a function of the distance from the galactic center, are inherently Newtonian constructs. Their interpretation assumes the existence of global inertial frames and relies on Galilean addition of velocities to convert Doppler shifts measured within the Solar System into velocities relative to the galactic center~\cite{SofueRubin2001}. Such procedures break down in curved spacetime, where tangent spaces at different events cannot be directly compared due to the absence of a global notion of parallelism. 
Moreover, the interpretation of the observed Doppler shifts in terms of physical radial velocities is model-dependent and in general requires a full description of the observational process~\cite{LindegrenDravins2003}.  A consistent general relativistic formulation must therefore redefine rotation curves in terms of locally measurable quantities, taking into account that all astronomical observations involve the propagation of light. These include spectroscopic and astrometric relative velocities~\cite{2007CMaPh.273..217B}, which shall be evaluated and compared within locally inertial frames, providing a coherent framework for interpreting observational data in a fully general relativistic setting.

The paper is organized as follows. In Section~\ref{Stationary axisymmetric spacetimes} we introduce the general form of a stationary, axisymmetric spacetime coupled to dust, restricting to the particular case of rigidly rotating dust in ~\ref{BG model}. Section~\ref{Reference frames and relative velocities} is devoted to reference systems and relative velocities. In particular,~\ref{reference systems} outlines how reference systems are constructed in GR, while~\ref{relative velocities} defines the relevant notions of relative velocity. The BCRS and ZAMOs are introduced and analyzed in~\ref{BCRS} and~\ref{ZAMO}, where their interpretation of a self-gravitating disc of dust is discussed. Section~\ref{Locally inertial frame} presents a general procedure for constructing a locally inertial reference system, which is then implemented for the models under consideration in ~\ref{geodesics},~\ref{gyroscopes},~\ref{Fermi coordinates}, and~\ref{metric Fermi}. In Section~\ref{radially locked reference system} we introduce the radially locked reference system, in which one spatial axis of the locally inertial system is aligned with the radial direction. Section~\ref{frequency shift} analyzes the frequency shift measured by locally inertial observers, and Section~\ref{calculation of relative velocity} proposes a method for computing the spectroscopic and astrometric relative velocities of a dust disc with respect to a locally inertial observer. Finally, Section~\ref{conclusions} contains our conclusions, while additional mathematical details are provided in Appendices~\ref{A: Gyroscope equation solution},~\ref{A: curvature quantities}, and~\ref{A: Riemann locally inertial reference system}.

\paragraph{Notation.} We adopt the metric signature $(-+++)$ and work in geometrized units where $c=G=1$. The scalar product of two vectors $u$ and $v$ is denoted by $g(u,v)$ without referring to a particular coordinate system, or by $g_{\mu\nu}u^{\mu}v^{\nu}$ when working in component notation; the spatial norm of a vector \(v\) is denoted by \(\norm{v}\). The partial derivative of a scalar function $f$, \(\pdv{f}{x}\), is denoted by \(f_{,x}\).

%%%%%%%%%%%%%%%%%%%%%%%%%%%%%%%%%%%%%%%%

\section{Stationary axisymmetric self-gravitating dust}
\label{Stationary axisymmetric spacetimes}

We consider a disc of self-gravitating dust, modeled as a stationary and axisymmetric solution of the EE. Such spacetimes admit two Killing vector fields: a timelike vector $\xi$ associated with stationarity, and a spacelike vector $\psi$ associated with axial symmetry, whose Killing product vanishes. 
Physically, this expresses the intuitive fact that performing a rotation followed by a time translation yields the same result as performing them in the opposite order. Such a property characterizes isolated systems in equilibrium and has been proven to hold for asymptotically flat stationary axisymmetric spacetimes~\cite{Carter:1970ea}.
In addition, we assume that the circularity condition holds. Specifically, a spacetime is said to be circular if it admits $2$-spaces, called meridional surfaces, orthogonal to the orbits of the Killing vectors. According to Frobenius’ theorem, this geometric requirement can be reformulated as a condition on the Ricci tensor: a stationary axisymmetric spacetime admits such orthogonal $2$-spaces if and only if~\cite{KundtTrumper1966}
\begin{equation}
\label{circularity condition Ricci}
\xi^{a}R_{a}{}^{[b}\xi^{c}\psi^{d]}=\psi^{a}R_{a}{}^{[b}\xi^{c}\psi^{d]}=0.
\end{equation}
These conditions are fulfilled in most stationary axisymmetric spacetimes of physical interest. In particular, they hold for perfect fluid solutions provided that the four-velocity of the fluid $u$ satisfies
\begin{equation}
\label{circularity condition four-velocity}
    u_{[a}\xi_{b}\psi_{c]}=0.
\end{equation}
This is also called circularity condition for such matter configurations~\cite{Carter1969}. The equivalence between the geometric and physical forms of the circularity condition illustrates the close correspondence between matter and geometry in GR: roughly speaking, circular sources give rise to circular spacetimes. \\
For spacetimes satisfying the condition~\eqref{circularity condition Ricci}, one can introduce coordinates adapted to the foliation by transitivity and meridional surfaces, $(t,r,z,\phi)$, such that $\xi=\partial_{t}$ and $\psi=\partial_{\phi}$. In these coordinates, the metric takes the general Lewis–Papapetrou–Weyl (LPW) form~\cite{Wald:1984, Stephani:2003tm},
\begin{align}
\label{ZAMO metric}
 ds^2=-e^{2\nu} dt^2+e^{2\psi} (d\phi-\chi dt)^2+e^{\mu} (dr^2+dz^2),
\end{align}
where, by virtue of the Killing equations, all metric functions depend only on $r$ and $z$. \\
The stress-energy tensor of the dust is 
\begin{equation}
    T_{\mu\nu}=\rho u_{\mu}u_{\nu},
\end{equation}
where $\rho$ is the matter density. The circularity condition~\eqref{circularity condition four-velocity} constrains $u$ to lie in the surfaces of transitivity spanned by the two Killing vectors, so that it can be written as
\begin{equation}
\label{dust velocity}
    u=\frac{1}{\sqrt{-H}}(\partial_{t}+\Omega\partial_{\varphi}),
\end{equation}
where $\Omega=\frac{d\phi}{dt}$ is the dust angular velocity. The function $H(r,z)$ is determined by the normalization condition $u^{\mu}u_{\mu}=-1$, yielding
\begin{equation}
\label{expression for H}
    H=-e^{2\nu}+e^{2\psi}(\chi-\Omega)^{2}.
\end{equation}
Each Killing vector field gives rise to a conserved quantity along geodesics. Therefore, for a particle with four-velocity $u$~\eqref{dust velocity} moving in a stationary axisymmetric spacetime, there exist two conserved quantities: the specific energy and the specific angular momentum,
\begin{equation}
\label{conserved quantities}
    E:=-g_{\mu\nu}\xi^{\mu}u^{\nu}, \qquad
    L:=g_{\mu\nu}\psi^{\mu}u^{\nu}.
\end{equation}
The metric, as well as the dust density and four-velocity, can be conveniently expressed in terms of an arbitrary function $\eta(r,z)$ and a related negative function $H(\eta(r,z))$~\cite{Stephani:2003tm, Astesiano:2021ren}. 
More precisely,
\begin{subequations}
\begin{align}
    e^{2\nu}=&\frac {r^2}{g_{\phi\phi}}=-\frac {Hr^2}{r^2-\eta^2}, \\
    e^{2\psi}=&g_{\phi\phi}=-\frac {r^2-\eta^2}{H}, \\ \chi=&-\frac {g_{t\phi}}{g_{\phi\phi}} =\Omega+\frac {H\eta}{r^2-\eta^2} \\
    \mu,_{r}=&\frac{1}{2r}[g_{tt,r}g_{\phi\phi,r}-g_{tt,z}g_{\phi\phi,z}-(g_{t\phi,r})^{2}+(g_{t\phi,z})^{2}],
    \label{eq:mur differentially rotating}\\
    \mu,_{z}=&\frac{1}{2r}[g_{tt,z}g_{\phi\phi,r}-g_{tt,r}g_{\phi\phi,z}-2g_{t\phi,z}g_{t\phi,r}],
    \label{eq:muz differentially rotating}
\end{align}
\end{subequations}
where
\begin{equation}
\label{equation for omega}
    \Omega=\frac{1}{2}\int \frac{dH}{d\eta}\frac{d\eta}{\eta}
\end{equation}
gives the angular velocity of the dust. Correspondingly, the energy density of the dust is
\begin{equation}
\label{matter density}
    8\pi\rho=\frac{\eta^{2}r^{-2}(2-\eta l)^{2}-r^{2}l^{2}}{4g_{rr}}\frac{\eta,_{r}^{2}+\eta,_{z}^{2}}{\eta^{2}},
\end{equation}
with
\begin{equation}
    l=\frac{H'}{H}.
\end{equation}
Here a prime denotes differentiation with respect to $\eta$. \\
Finally, notice that $\eta$ and $H$ cannot be chosen independently, in the sense that for an arbitrary choice of $H(\eta)$, one can introduce the auxiliary function
\begin{equation}
\label{def F}
    \mathcal{F}=2\eta+r^{2}\int \frac{H'}{H}\frac{d\eta}{\eta}-\int\frac{H'}{H}\eta d\eta,
\end{equation}
which is constrained by the EE to satisfy
\begin{equation}
\label{F Grad-Shafranov}
    \mathcal{F},_{rr}-\frac{1}{r}\mathcal{F},_{r}+\mathcal{F},_{zz}=0.
\end{equation}
Replacing \eqref{def F} into~\eqref{F Grad-Shafranov}, one gets an equation for $\eta(r,z)$, 
\begin{equation}
\label{equation for eta}
    \left(\eta,_{ r r}-\frac{1}{r} \eta,_{ r}+\eta,_{ z z}\right)(2-\eta \ell)+\left(\eta,_{ r}^2-\eta,_{ z}^2\right)\left(\eta \ell^{\prime}-\ell\right)\left(1+\frac{r^2}{\eta^2}\right)+r^2 \frac{\ell}{\eta}\left(\eta,_{ r r}+\frac{3}{r} \eta,_{ r}+\eta,_{ z z}\right)=0.
\end{equation}
Given a function $H(\eta)$, it is then possible to solve the above equation for $\eta$ and thus specify all the metric quantities, as well as the dust density and four-velocity. Analytical solutions exist for rigidly rotating dust, corresponding to $l=0$ (see Section~\ref{BG model}), while for the general case of differentially rotating dust $(l\neq 0)$ only approximate solutions have been proposed~\cite{Astesiano:2021ren}.

%%%%%%%%%%%%%%%%%%%

\subsection{Rigidly rotating self-gravitating dust}
\label{BG model}

\begin{figure}
\centering
\includegraphics[scale=0.55]{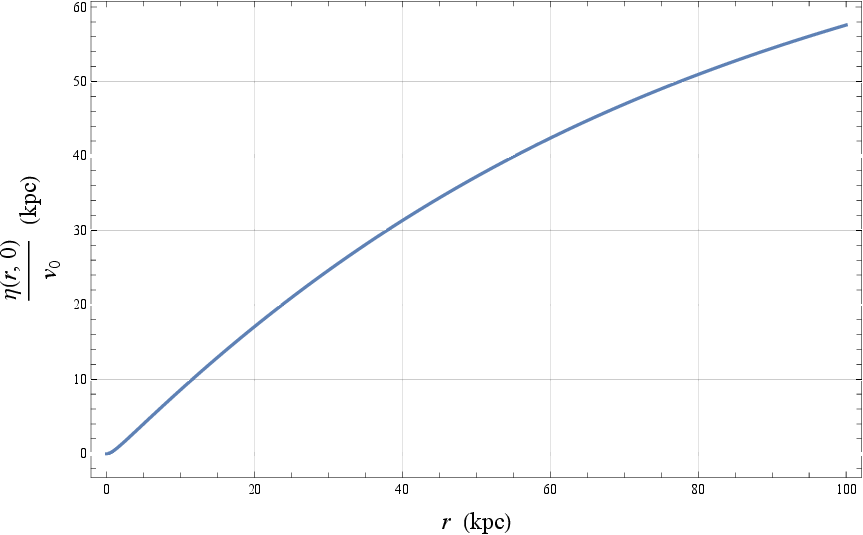}
\caption{$\eta(r,0)$ as a function of $r$. The parameters values are: $v_{0} = \SI{200}{\kilo\metre\per\second} \simeq 7\times 10^{-4}$, $r_{0} = \SI{1}{\kilo\parsec}$, $R = \SI{100}{\kilo\parsec}$, as in \cite{BG}.}
\label{eta_BG_galactic_plane}
\end{figure}

We now consider the case of rigidly rotating dust, characterized by a constant angular velocity $\Omega=constant$~\cite{VanStockum1938,Bonnor1980}. By performing a rigid rotation of the coordinate system, one can adopt the comoving frame in which $\Omega=0$, so that the dust four-velocity becomes
\begin{equation}
    u=\xi=\partial_{t}.
\end{equation}
This follows directly from the general expression of the four-velocity~\eqref{dust velocity} obtained from the circularity condition. The EE imply that imposing a constant angular velocity also forces the function $H$ to be constant, as can be seen directly from Eq.~\eqref{equation for omega}. By a suitable rescaling of the time coordinate, one may then set $H=-1$. As a result, the metric~\eqref{ZAMO metric} simplifies to
\begin{equation}
\label{rigidly rotating dust metric}
    ds^{2}=-(dt-\eta d\phi)^{2}+r^{2}d\phi^{2}+e^{\mu}(dr^{2}+dz^{2}).
\end{equation}
From eq.~\eqref{def F} we also have $\mathcal{F}=2\eta$, so that eq.~\eqref{F Grad-Shafranov} reduces to the Grad-Shafranov equation for $\eta$:
\begin{equation}
\label{eq: eta rigid}
    \eta,_{rr}-\frac{1}{r}\eta,_{r}+\eta,_{zz}=0.
\end{equation}
The equations~\eqref{eq:mur differentially rotating} and~\eqref{eq:muz differentially rotating} which determine the function $\mu$, and the equation \eqref{matter density} for the dust matter density take the form
\begin{subequations}
\begin{align}
    &\mu,_{r}=-\frac{1}{2r}(\eta,_{r}^{2}-\eta,_{z}^{2}), \label{eq: mur rigid} \\
    &\mu,_{z}=-\frac{1}{r}\eta,_{r}\eta,_{z}, \label{eq: muz rigid} \\
    &8\pi \rho=\frac{e^{-\mu}}{r^{2}}(\eta,_{r}^{2}+\eta,_{z}^{2}). \label{eq: matter density rigid}
\end{align}
\end{subequations}
It follows that any solution of~\eqref{eq: eta rigid} completely specifies a class of rigidly rotating dust solutions of the EE for stationary axisymmetric spacetimes. Among these, we seek a solution that can effectively describe a dust disc within an appropriate regime of validity. \\
To solve~\eqref{eq: eta rigid} we apply a spectral separation of variables and impose the physical boundary conditions $\eta(r,z)\to 0$ for $r\to\infty$ and $|z|\to\infty$,
\begin{align}
 \eta(r,z)=\int_0^\infty \mathcal{R}(r,\lambda)\mathcal{Z}(z,\lambda) d\lambda.
\end{align}
This approach leads to
\begin{equation}
\label{eq: generic solution for eta in and out the galaxy plane}
    \eta(r,z)=\frac{r^{2}}{2}\int_{0}^{\infty} C(x)\Big([(z+x)^{2}+r^{2}]^{-\frac{3}{2}}+[(z-x)^{2}+r^{2}]^{-\frac{3}{2}}\Big) dx,
\end{equation}
which is expressed in terms of an arbitrary spectral function $C(x)$. 
In the spirit of~\cite{BG}, we impose the following, less restrictive, conditions:
\begin{itemize}
    \item $v_{Z}(r,0)\propto r$ for small values of $r$;
    \item $v_{Z}(r,0)\propto \frac{1}{r}$ asymptotically, well outside the disc,
\end{itemize}
where $v_{Z}$ is the rotation velocity as measured by Zero Angular Momentum Observers; see Section~\ref{ZAMO}. These boundary conditions single out a specific rigidly rotating dust configuration, known as Balasin-Grumiller (BG) model \cite{BG}, whose velocity profile, relative to this class of observers, reproduces the main observed features of galactic rotation curves~\cite{Crosta:2020,Beordo:2024}. Different choices of boundary conditions would correspond to solutions with distinct phenomenological behavior. \\
The simplest spectral function that provides a velocity profile satisfying the above conditions is 
\begin{equation}
\label{BG spectral density}
    C(x)=(x-r_{0})v_{0}[\theta(x-r_{0})-\theta(x-R)]+(R-r_{0})\theta(x-R),
\end{equation}
where $\theta(x)$ is the Heaviside step function, $v_{0}$ is a constant velocity (relative to the speed of light), $r_{0}$ is an internal radius, and $R$ is the radius of the disc of dust~\cite{BG}. With this choice, Eq.~\eqref{eq: generic solution for eta in and out the galaxy plane} becomes
\begin{equation}
\label{eq: simplified expression of eta for the particular choice of spectral density}
    \eta(r,z)=v_{0}(R-r_{0})+\frac{v_{0}}{2}\left[\sqrt{(z+r_{0})^{2}+r^{2}}+\sqrt{(z-r_{0})^{2}+r^{2}}-\sqrt{(z+R)^{2}+r^{2}}-\sqrt{(z-R)^{2}+r^{2}}\right].
\end{equation}
It is important to emphasize that this solution has a limited range of applicability within the dust approximation, which requires sufficiently small departures from the galactic plane. In particular, for $|z|>r_0$, it occurs that $r^{2}\leq \eta^{2}$, causing the $\phi$ coordinate to become timelike. In such cases, the orbits of the Killing vector $\partial_{\phi}$ (circles of constant $r$ and $z$) become closed timelike curves; therefore, to avoid this issue, we will restrict the analysis to the galactic plane $z=0$, on which the function $\eta$ reads
\begin{equation}
\label{eta BG galactic plane}
    \eta(r,0)=v_{0}(R-r_{0})+v_{0}\left[\sqrt{r_{0}^{2}+r^{2}}-\sqrt{R^{2}+r^{2}}\right].
\end{equation}
A plot of the function $\eta(r,0)$ is shown in Figure~\ref{eta_BG_galactic_plane}, evaluated with the parameter values adopted in \cite{BG}. 

\begin{figure}
\centering
\includegraphics[scale=0.55]{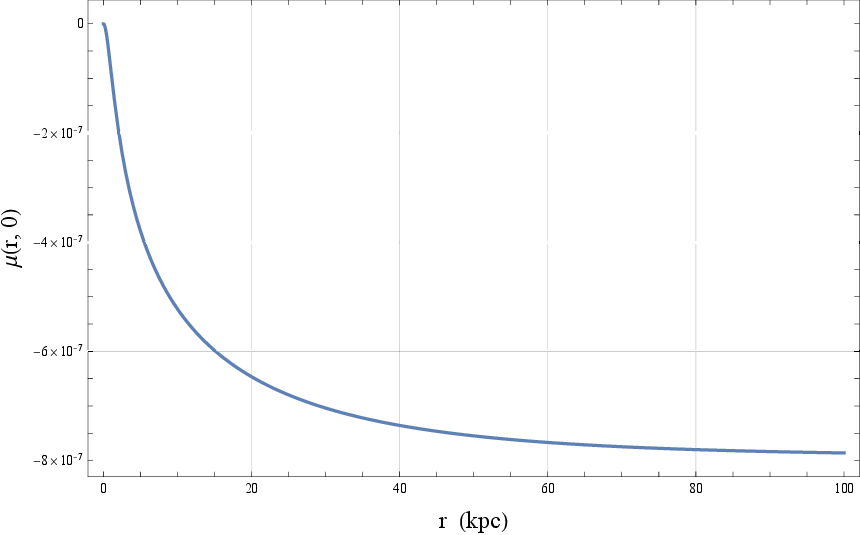}
\caption{$\mu(r,0)$ as a function of $r$. The parameters values are: $v_{0} = \SI{200}{\kilo\metre\per\second} \simeq 7\times 10^{-4}$, $r_{0} = \SI{1}{\kilo\parsec}$, $R = \SI{100}{\kilo\parsec}$, as in \cite{BG}.}
\label{mu_BG_galactic_plane}
\end{figure}

Restricting to the galactic plane and assuming reflection symmetry with respect to it, Eqs.~\eqref{eq: mur rigid} and~\eqref{eq: muz rigid} reduce to 
\begin{equation}
    \mu,_{z}=0, \qquad
    \mu,_{r}=-\frac{\eta,_{r}^{2}}{2r}.
\end{equation}
The first is trivial, while the second can be integrated using the expression~\eqref{eta BG galactic plane} for $\eta$, 
\begin{align}
\label{eq: explicit expression for mu}
    \mu(r,0)=-\frac{v_0^2}{2}&\left[\frac{1}{2}\ln(r^2+r_0^2)+\frac{1}{2}\ln(r^2+R^2) \right. \notag \\
    &\left.-\ln\left(\sqrt{r^{2}+r_{0}^{2}}+\sqrt{r^{2}+R^{2}}\right)^{2}\right]+ \frac{v_0^2}{2}\ln(\frac{Rr_0}{(R+r_0)^2})+ \mu_0, 
\end{align}
where $\mu_0=\mu(0,0)$ is an integration constant. A plot is shown in Figure~\ref{mu_BG_galactic_plane}. 

\begin{figure}[htbp]
    \centering
    \begin{minipage}{0.48\textwidth}
        \centering
        \includegraphics[width=\linewidth]{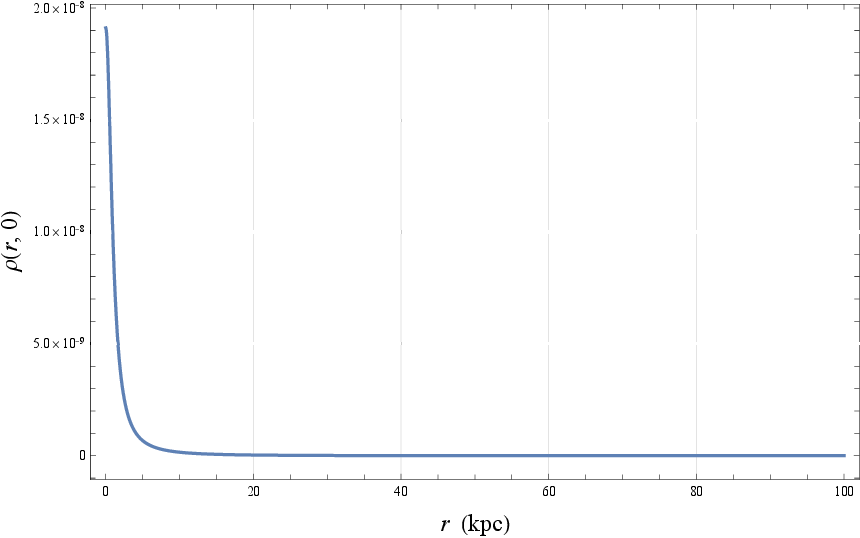}
    \end{minipage}
    \hfill
    \begin{minipage}{0.48\textwidth}
        \centering
        \includegraphics[width=\linewidth]{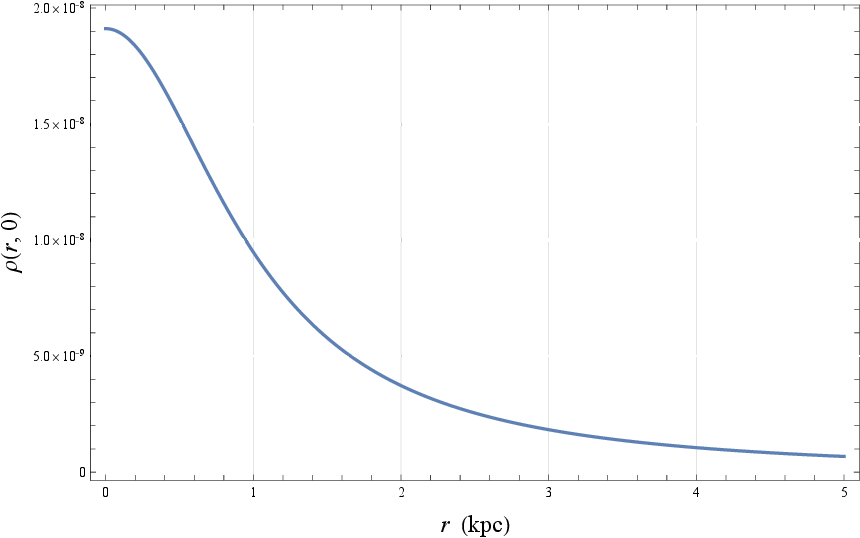}
    \end{minipage}
    \caption{Left panel: $\rho(r,0)$ as a function of $r$. Right panel: zoomed view near the origin. The parameters values are: $v_{0} = \SI{200}{\kilo\metre\per\second} \simeq 7\times 10^{-4}$, $r_{0} = \SI{1}{\kilo\parsec}$, $R = \SI{100}{\kilo\parsec}$, as in \cite{BG}.}
\label{rho_BG_galactic_plane}
\end{figure}

The function $\mu(r)$ determines the geometry of the galactic plane ($z=0$). Setting $\mu_0=0$, it is possible to remove the conical singularity along the rotation axis, but this does not eliminate all global angular defects: a conical structure reappears asymptotically, far from the axis. This feature is intrinsic to the geometry and cannot be removed by any choice of boundary conditions. It represents one of the genuinely non-Newtonian aspects of general relativistic self-gravitating dust configurations, absent in the Newtonian approximation~\cite{Fontana:2024aru}. \\
We can then complete the characterization of the model by computing the dust matter density on the galactic plane from Eq.~\eqref{eq: matter density rigid}, by direct substitutions of Eqs.~\eqref{eta BG galactic plane} and~\eqref{eq: explicit expression for mu}:
\begin{equation}
    \rho(r,0)=\frac{v_{0}^{2}}{8\pi}e^{-\mu_{0}}\left[\frac{(R+r_{0})^{2}\sqrt{r^{2}+r_{0}^{2}}\sqrt{r^{2}+R^{2}}}{r_{0}R\left(\sqrt{r^{2}+r_{0}^{2}}+\sqrt{r^{2}+R^{2}}\right)^{2}}\right]^{\frac{v_{0}^{2}}{2}}\left(\frac{1}{\sqrt{r^{2}+R^{2}}}-\frac{1}{\sqrt{r^{2}+r_{0}^{2}}}\right)^{2}.
\end{equation}
A plot is shown in Figure~\ref{rho_BG_galactic_plane}.

For $r\ll r_{0}$, $\rho\simeq\dfrac{e^{-\mu_{0}}v_{0}^{2}}{8\pi}\dfrac{(R-r_{0})^{2}}{R^{2}r_{0}^{2}}+\mathcal{O}(r^{2})$, approaching a constant value at the center; for $r_{0}\ll r\ll R$, $\rho\propto r^{-2+\frac{v_{0}^{2}}{2}}$; and for $r\gg R$, $\rho\propto r^{-6}$, showing an asymptotic power-law decay. \\
Some comments on the physical viability of the BG solution are in order. The model should be regarded as a simplified toy model, as it relies on several idealizations. Like other exact solutions of the EE used to describe galactic discs, it assumes stationarity and axial symmetry, and treats the stellar component as a continuous, pressureless perfect fluid, an approximation valid only for the outer stellar disc, beyond the central bulge. Moreover, the assumption of rigid rotation is unphysical, since stars at different radii are observed to rotate with distinct angular velocities. 
Despite these limitations, the BG model is simple enough to allow for a fully analytic treatment while still capturing essential general-relativistic features of self-gravitating galactic configurations, which persist in more realistic, differentially rotating models.

%%%%%%%%%%%%%%%%%%%%%%%%%%%%%%%%%%%%%

\section{Observers and relative velocities}
\label{Reference frames and relative velocities}

The physical interpretation of exact solutions to the EE is notoriously challenging, in part because of the difficulty of establishing suitable coordinate systems in GR. In Newtonian mechanics, all clocks measure the same absolute time, and space is treated as a fixed Euclidean background. In this framework, it is possible to define global inertial observers related by Galilean transformations, involving constant translations, rotations, and uniform velocities, all providing the same physical description of the process under consideration. They share the same notions of time, distance, and simultaneity, and free particles move along straight lines indefinitely. This global inertial structure forms the basis of the Newtonian formulation of dynamics. \\
In GR, on the contrary, it is not possible to define globally inertial observers due to the spacetime curvature. The best one can do is to set up locally inertial reference systems covering a limited portion of the spacetime. Astronomical practice, by contrast, still relies on the existence of quasi-inertial frames based on a post-Newtonian approximation, a prominent example being the Barycentric Celestial Reference System (BCRS)~\cite{2005USNOC.179.....K, Soffel_2003}. \\
In this section, we first review the notion of reference systems in GR, with particular emphasis on the BCRS and Zero Angular Momentum Observers, and the various possible definitions of relative velocity. We then present the construction of a locally inertial reference system adapted to an observer following a dust geodesic in a $(\eta,H)$ spacetime, intended to model the averaged dynamics of a galactic disc. Finally, we propose a procedure to define a shared spatial direction across the different local patches covering the disc, without relying on the existence of distant non-rotating objects, by identifying radial photons passing through the galactic center.

%%%%%%%%%%%%%%%%%%%%%%%%%%%%%%%%%%%%%

\subsection{Reference systems}
\label{reference systems}

A free-falling observer is represented by a timelike geodesic $\Gamma$ with four-velocity field $u$. At each event \( p \in \Gamma \), the tangent space \( T_{p}\mathcal{M} \) splits into the temporal direction spanned by \( u \) and the local rest space $u^{\perp} = \{\, v \in T_{p}\mathcal{M} \, | \, g(u,v) = 0 \,\},$ which is the three-dimensional subspace containing all the vectors orthogonal to $u$. Any vector lying in $u^{\bot}$ is purely spatial from the point of view of $u$. Within its local rest space, the observer introduces a set of spatial axes $e_{i}$, whose transport law along the observer's worldline is in principle arbitrary. Together with the four-velocity, this construction yields a tetrad \{${e_{0}=u,e_{i}}$\} which can be used to measure tensor components. \\
To locate events in spacetime, a coordinate system \(x^{\alpha}\) = \{\(x^{0}\), \(x^{i}\)\} is required. Such systems arise naturally from the construction outlined above: when the basis vector $\partial_{0}$ is timelike, it is tangent to an observer's worldline, while the projections of the spacelike coordinates basis vectors $\partial_{i}$ onto the observer's rest space provide a system of spatial axes in $u^{\perp}$. A coordinate system then serves to label spacetime events, with $x^{i}$ specifying the position relative to the spatial axes and $x^{0}$ indicating the corresponding time along the observer's worldline. \\
In the context of relativistic astronomy, any observation involves four essential elements: the motion of the observer along its worldline, the motion of the observed object, the propagation of light between them, and the act of measurement itself. Each of these components must be described consistently within the relativistic framework. The reference system provides the theoretical and operational conventions\footnote{It is useful to draw here a distinction and appreciate the difference between reference \emph{system} and \emph{frame}. While the first refers to the theoretical definition for a system of coordinates, the second relates to its practical implementation, specifying e.g. a procedure to assign and calculate coordinates to physical points/events.} needed to assign coordinates to physical events, ensuring a coherent treatment of light propagation, motion of celestial bodies, and time synchronization~\cite{SoffelLanghans2013}.

%%%%%%%%%%%%%%%%%%%%%%%%%%%%%%%%%%%%

\subsection{Relative velocities}
\label{relative velocities}

Once an observer and its adapted reference system have been defined, one must specify how to measure the relative velocity of another observer or particle with respect to it. In Newtonian mechanics, this notion is unambiguous: all velocities can be compared directly because they are defined in a common Euclidean space and share the same absolute time. In GR, however, the situation is more subtle. The four-velocities of observers at different spacetime points belong to distinct tangent spaces and therefore cannot be compared directly. A well-defined notion of relative velocity exists only when the two observers coincide at the same event~\cite{2007CMaPh.273..217B}.

\paragraph{Observers at the same event.} 
Given two observers $u$, $u'$ at the same event $p$ there exists a unique vector $v\in u^{\bot}$ (rest space of $u$) and a unique real number $\gamma$ such that 
\begin{equation}
\label{relative velocity same event}
    u'=\gamma(u+v).
\end{equation}
As a consequence, $0\leq ||v||<1$ and $\gamma=-g(u',u)=\frac{1}{\sqrt{1-||v||^{2}}}$. We say that $v$ is the relative velocity of $u'$ observed by $u$ and $\gamma$ is the gamma factor corresponding to the velocity $||v||$. It is possible to obtain $||v||$ directly from the gamma factor,
\begin{equation}
\label{velocity gamma}
    ||v||=\sqrt{\frac{\gamma^{2}-1}{\gamma^{2}}}.
\end{equation}
Intuitively, an observer moving with four-velocity $u$ would measure $v$ as the special relativistic three-velocity of $u'$ in its local inertial frame because $v$ is a purely spatial vector from the point of view of $u$ since it belongs to its rest space.

\begin{figure}
\centering
\includegraphics[scale=0.8]{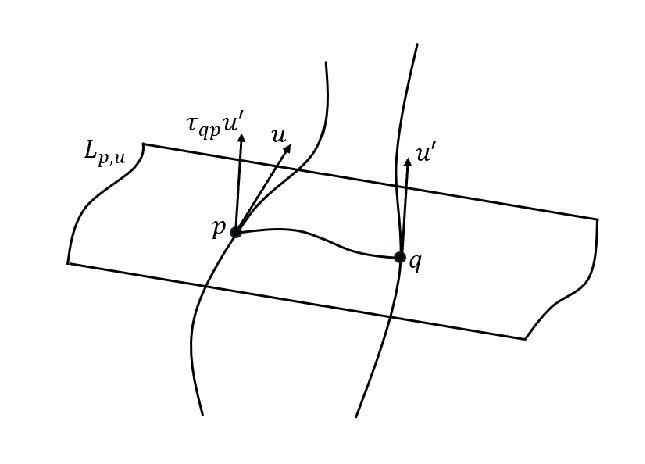}
\caption{Schematic illustration of the definition of the kinematic relative velocity (adapted from~\cite{2007CMaPh.273..217B}).}
\label{fig: kinematic relative velocity}
\end{figure}

\paragraph{Kinematic relative velocity.} 
As already remarked, if the two observers are at different events, it is not possible to uniquely define their relative velocity. A straightforward generalization of the concept of relative velocity between two observers at different events is the kinematic relative velocity, which rests on the concept of spacelike simultaneity~\cite{BOLOS2006813,2007CMaPh.273..217B}. In particular, considering an observer $u$ at $p\in\mathcal{M}$ we define $\varphi:\mathcal{M}\to \mathbb{R}$ such that $\varphi(q):=g(\exp_{p}^{-1}q,u)$, where $\exp_{p}:T_{p}\mathcal{M}\to\mathcal{M}$\footnote{Let $(\mathcal{M},g)$ be a smooth manifold, let $\nabla$ be the corresponding Levi-Civita connection, and $p\in\mathcal{M}$. For any vector $v\in T_{p}\mathcal{M}$ consider the unique geodesic $\gamma_{v}:I\to\mathcal{M}$ satisfying $\gamma_{v}(0)=p$, $\dot{\gamma}(0)=v$, where $I\in\mathbb{R}$ is a maximal interval of existence, with $[0,1]\subseteq I$. The exponential map at $p$ is the map $exp_{p}:T_{p}M\to \mathcal{M}$ defined by $exp_{p}(v)=\gamma_{v}(1)$.} is the exponential map. The set $L_{p,u}:=\varphi^{-1}(0)$ is a regular three-dimensional submanifold, called the Fermi surface. Intuitively, $L_{p,u}$ is the subset of $\mathcal{M}$ containing the points that are connected to $p$ by a spacelike geodesic with tangent vector in $u^{\bot}$. In this sense, an event in such space is spacelike simultaneous with $p$. Physically, the events in $L_{p,u}$ cannot influence the observer at $p$, since any causal interaction, whether mediated by electromagnetic or gravitational waves, propagates at the speed of light. \\
Consider now two observers $u$, $u'$ at $p$, $q$, respectively, such that $q\in L_{p,u}$. The kinematic relative velocity of $u'$ with respect to $u$ is the unique vector $v_{kin}\in u^{\bot}$ such that 
\begin{equation}
    \tau_{qp}u'=\gamma(u+v_{kin}),
\end{equation}
where $\tau_{qp}$ is the parallel transport operator from $q$ to $p$, and $\gamma$ is the gamma factor corresponding to the velocity $||v_{kin}||$. Basically, we consider an observer $u$ at $p$ and an observer $u'$ at $q$, with $p$ and $q$ spacelike simultaneous, which means that there exists a spacelike geodesic from $q$ to $p$ with tangent vector in the rest space of $u$. Then, we parallel transport $u'$ from $q$ to $p$ along the spacelike geodesic connecting these two points. In this way, we obtain the vector $\tau_{qp}u'$ which belongs to the same tangent space of $u$ ($T_{p}\mathcal{M}$). The kinematic relative velocity of $u'$ with respect to $u$ is then the standard relative velocity of $\tau_{qp}u'$ with respect to $u$. A schematic representation of this construction is shown in Fig.~\ref{fig: kinematic relative velocity}.

\begin{figure}
\centering
\includegraphics[scale=0.8]{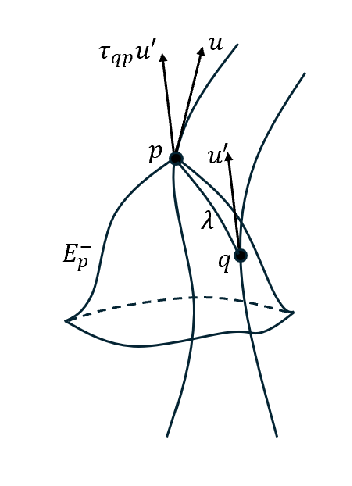}
\caption{Schematic illustration of the definition of the spectroscopic relative velocity (adapted from~\cite{2007CMaPh.273..217B}).}
\label{fig: spectroscopic relative velocity}
\end{figure}

\paragraph{Spectroscopic relative velocity.}
The measurements of the velocities of celestial bodies is made through light rays; it involves events connected by null geodesics, meaning that they are lightlike simultaneous. Therefore, it makes sense to seek for definitions of relative velocities based on lightlike simultaneity that can be related to physical observables, such as the spectroscopic relative velocity~\cite{BOLOS2006813, 2007CMaPh.273..217B}. \\
A light ray is represented by a lightlike geodesic $\lambda$ and a future-pointing null vector field $F$ tangent to $\lambda$ and parallel transported along it, i.e., $\nabla_{F}F=0$. Given $p\in\lambda$ and $u$ an observer at $p$, there exists a unique vector $w\in u^{\bot}$ and a unique real positive number $\nu$ such that 
\begin{equation}
\label{eq: frequency vector}
    F_{p}=\nu(u+w).
\end{equation}
As consequences we have $||w||=1$ and $\nu=-g(F_{p},u)$. We say that $w$ is the relative velocity of $\lambda$ observed by $u$, and $\nu$ is the frequency of $\lambda$ observed by $u$. Before introducing the spectroscopic relative velocity, there are two other important concepts that need to be defined. Let $p\in\mathcal{M}$ and $\varphi:\mathcal{M}\to \mathbb{R}$ defined by $\varphi(q):=g(\exp_{p}^{-1}q,\exp_{p}^{-1}q)$. Then the set $E_{p}:=\varphi^{-1}(0)-{p}$ is a regular three-dimensional submanifold called horismos submanifold of $p$. An event $q$ is in $E_{p}$ if and only if $q\neq p$ and there exists a lightlike geodesic connecting $p$ and $q$. Basically, it is the preimage of the lightlike cone of $p$ through the exponential map. $E_{p}$ has two connected components, $E_{p}^{-}$ and $E_{p}^{+}$. $E_{p}^{-}$ ($E_{p}^{+}$) is the past-pointing (future-pointing) horismos submanifold of $p$ and it is the connected component of $E_{p}$ in which for each $q\in E_{p}^{-}$ ($E_{p}^{+}$) the preimage $\exp_{p}^{-1}q$ is a past-pointing (future-pointing) null vector. \\
Then, given $u$ an observer at $p$ and an observed event $q\in E_{p}^{-}\bigcup{p}$, the relative position of $q$ observed by $u$ is the projection of $\exp_{p}^{-1}q$ onto $u^{\bot}$, i.e.,
\begin{equation}
    s_{obs}:=P_u\exp_{p}^{-1}q=
    \exp_{p}^{-1}q+g(\exp_{p}^{-1}q,u)u,
\end{equation}
where $P_u=g+u\otimes u$ projects onto $u^{\bot}$. This corresponds to the line-of-sight direction. \\
We now have all the elements to characterize the spectroscopic relative velocity. Let $u$, $u'$ be two observers at $p$, $q$ such that $q\in E_{p}^{-}$ and let $\lambda$ be a light ray from $q$ to $p$. The spectroscopic relative velocity of $u'$ with respect to $u$ is the unique vector $v_{spec}\in u^{\bot}$ such that
\begin{equation}
\label{eq: spectroscopic relative velocity}
    \tau_{qp}u'=\gamma (u+v_{spec}),
\end{equation}
where $\tau_{qp}$ is the parallel transport operator from $q$ to $p$, and $\gamma$ is the gamma factor corresponding to the velocity $||v_{spec}||$. A schematic representation of this construction is shown in Fig.~\ref{fig: spectroscopic relative velocity}.

Intuitively, we consider two events $p$ and $q$ that are lightlike simultaneous, which means that there exists a light ray from $q$ to $p$. Then, consider an observer $u$ at $p$ and an observer $u'$ at $q$. To compute the spectroscopic velocity of $u'$ with respect to $u$, we have to parallel transport $u'$ from the ``emission event'' $q$ to the ``observation event'' $p$ along the null geodesic connecting these two points. In this way, we obtain the vector $\tau_{qp}u'$ which belongs to the same tangent space of $u$ ($T_{p}\mathcal{M}$). The spectroscopic relative velocity of $u'$ with respect to $u$ is then the relative velocity of $\tau_{qp}u'$ with respect to $u$. This can in turn be decomposed into a line-of-sight (radial) component and a tangential component. In particular, if $s_{obs}$ is the relative position of $q$ with respect to $u$ at $p$, the spectroscopic radial velocity of $u'$ with respect to $u$ is the component of $v_{spec}$ parallel to $s_{obs}$, i.e.,
\begin{equation}
    v_{spec}^{rad}:=g\left(v_{spec},\frac{s_{obs}}{||s_{obs}||}\right)\frac{s_{obs}}{||s_{obs}||}.
\end{equation}
The spectroscopic tangential velocity of $u'$ with respect to $u$ is the component of $v_{spec}$ orthogonal to $s_{obs}$,
\begin{equation}
    v_{spec}^{tan}:=v_{spec}-v_{spec}^{rad}.
\end{equation}
The importance of this definition lies in its direct connection with measurable frequency shifts. \\
Let $\lambda$ be a light ray from $q$ to $p$ and let $u$, $u'$ be two observers at $p$, $q$, respectively. Then,
\begin{equation}
\label{frequencies}
    \nu'=\gamma(1-g(v_{spec},w))\nu,
\end{equation}
where $\nu'$, $\nu$ are the frequencies of $\lambda$ observed by $u$, $u'$, respectively, $v_{spec}$ is the spectroscopic relative velocity of $u'$ observed by $u$, $w$ is the relative velocity of $\lambda$ observed by $u$, and $\gamma$ is the gamma factor corresponding to $||v_{spec}||$. This can be proven by recalling that parallel transport preserves the metric, so that
\begin{equation}
    \nu'=-g(F_{q},u')=-g(\tau_{qp}F_{q},\tau_{qp}u')=-g(F_{p},\tau_{qp}u'),
\end{equation}
where the last step follows from the fact that $F$ is parallel transported along $\lambda$. Substituting the expressions \eqref{eq: frequency vector} and \eqref{eq: spectroscopic relative velocity} one obtains 
\begin{equation}
    \nu'=-g[\nu(u+w),\gamma(u+v_{spec})].
\end{equation}
The result \eqref{frequencies} then follows from the fact that $g(u,u)=-1$, $g(u,w)=g(u,v_{spec})=0$ since $w,\,v_{spec}\in u^{\bot}$. \\
Eq.~\eqref{frequencies} is the general expression of the Doppler shift that includes gravitational redshift. Note that that $w=-\frac{s_{obs}}{||s_{obs}||}$, where $s_{obs}$ is the relative position of $q$ observed by $u$. Therefore, the spectroscopic radial velocity of $u'$ with respect to $u$ can be also written as 
\begin{equation}
    v_{spec}^{rad}=g(v_{spec},w)w.
\end{equation}
Taking the norm of this vector, one obtains
\begin{equation}
    ||v_{spec}||^{2}=g(v_{spec},w)^{2},
\end{equation}
which substituted in \eqref{frequencies} yields the familiar formula
\begin{equation}
\label{velocity from redshift}
    \frac{\nu'}{\nu}=\frac{1\pm ||v_{spec}^{rad}||}{\sqrt{1-||v_{spec}||^{2}}}.
\end{equation}
It is important to notice that it is not possible to obtain $v_{spec}$ directly from the frequency shift unless one of the two components of $v_{spec}$ is negligible with respect to the other. \\
The spectroscopic relative velocity is neither symmetric nor transitive. Consequently, it is not well-suited to obtain a galaxy rotation curve in the Newtonian sense, where Galilean velocity addition rules are used to compute velocities relative to the center of a galaxy from measurements made in the Solar System. Nevertheless, in Section~\ref{calculation of relative velocity}, we propose a possible general relativistic version of such concepts based on the construction of a locally inertial reference system. 

\paragraph{Astrometric relative velocity.}
To that end, we introduce yet another definition of relative velocity based on lightlike simultaneity, namely the astrometric relative velocity~\cite{BOLOS2006813,2007CMaPh.273..217B}. Let $u$, $u'$ be two observers at $p$, $q$ such that $q\in E_{p}^{-}$ and let $s_{obs}$ be the relative position of $q$ observed by $u$. The astrometric relative velocity of $u'$ with respect to $u$ is
the projection of $\nabla_{u}s_{obs}$ onto $u^{\bot}$, i.e.,
\begin{equation}
    v_{ast}:=\nabla_{u}s_{obs}+g(\nabla_{u}s_{obs},u)u.
\end{equation}
Intuitively, it gives the four-velocity of $u'$ at $q$ relative to $u$ at $p$ by measuring the variation of the relative position of $q$ with respect to $u$ in its proper time and projecting it in its rest space.
As usual, the astrometric relative velocity can be decomposed into a component parallel to $s_{obs}$ (the astrometric radial velocity) and one orthogonal to $s_{obs}$ (the astrometric tangential velocity). \\
The various definitions of relative velocity are summarized in Table~\ref{tab:relative_velocities}.

\begin{table}[h!]
\centering
\renewcommand{\arraystretch}{1.15}
\setlength{\tabcolsep}{5pt}
\begin{tabular}{|p{3cm}|p{4.8cm}|p{5.3cm}|}
\hline
\textbf{Type} & \textbf{Definition} & \textbf{Meaning / Properties} \\ 
\hline
\textbf{Same event} &
$u' = \gamma (u + v)$, $v \in u^{\perp}$ &
Standard local definition; coincident observers. \\ \hline

\textbf{Kinematic} &
\makecell[l]{
$\tau_{qp}u' = \gamma (u + v_{\mathrm{kin}})$ \\[2pt]
$q \in L_{p,u}$
} &
Based on spacelike simultaneity. \\ \hline

\textbf{Spectroscopic} &
\makecell[l]{
$\tau_{qp}u' = \gamma (u + v_{\mathrm{spec}})$ \\[2pt]
$q \in E_{p}^{-}$
} &
Defined along null geodesics; related to redshift. \\ \hline

\textbf{Astrometric} &
$v_{\mathrm{ast}} = \nabla_{u}s_{\mathrm{obs}}+g(\nabla_{u}s_{\mathrm{obs}},u)u$ &
From apparent position; describes proper motion. \\ \hline
\end{tabular}
\caption[Summary of relative velocities]{Summary of the main definitions of relative velocity in GR and their physical interpretation.}
\label{tab:relative_velocities}
\end{table}

%%%%%%%%%%%%%%%%%%%%%%%%%%%%%%%%%%%%%%%%%%%

\subsection{The BCRS}
\label{BCRS}
The construction of a reference system suitable for astronomical observations made on Earth or by near-Earth satellites within the framework of GR requires a precise measurement model. Earlier studies~\cite{1981CeMec..23...33M,1981CeMec..23...57M,AshbyBertotti1984,AshbyBertotti1986,1990CeMDA..48..167H} developed both barycentric and geocentric inertial and non-inertial reference systems, along with the transformations between them, within the post-Newtonian approximation. This effort ultimately was integrated into the formulation of the Barycentric Celestial Reference System (BCRS) and of the Geocentric Celestial Reference System (GCRS)~\cite{Soffel_2003,2005USNOC.179.....K,SoffelLanghans2013}. Adopted by the International Astronomical Union (IAU) in their 2000 resolutions, the BCRS provides a global framework for describing the positions and motions of Solar System bodies as well as galactic and extragalactic sources. The IAU resolutions prescribe the form of the metric tensor for both the barycentric and geocentric reference systems, and specify the coordinate transformations that relate them. More precisely, the BCRS is defined within GR by idealizing the Solar System as an isolated gravitating system, that is, neglecting all matter and fields external to it and assuming that the metric approaches the Minkowski form at infinity\footnote{Relaxing this hypothesis, a series of issues arise, e.g. the degree at which the cosmological expansion influences local dynamics and kinematics~\cite{2010RvMP...82..169C}.}. While this assumption is adequate for most Solar System applications, it is an idealization from the standpoint of cosmology and galactic dynamics: the mass-energy distribution of the Galaxy and other extragalactic sources, which introduce frame dragging and non-linear contributions, is ignored. Such effects become essential in the context of exact general relativistic galactic models, where non-Newtonian features of the metric cannot be treated as small perturbations. \\
The BCRS employs coordinates $(t,x^{i})$, where $t$ is the barycentric coordinate time. The metric tensor is given by a post-Newtonian approximation,
\begin{subequations}
\begin{align}
    &g_{00}=-1+2w^{0}-2(w^{0})^{2}+\mathcal{O}(6), \\
    &g_{0i}=-4w^{i}+\mathcal{O}(5), \\
    &g_{ij}=\delta_{ij}\left(1+2w^{0}\right)+\mathcal{O}(4),
\end{align}
\end{subequations}
where $w^{0}$ is a scalar potential generalizing the Newtonian gravitational potential, and $w^{i}$ is a vector potential associated with mass currents,  corresponding to the standard gravitomagnetic potential. Here, $\mathcal{O}(n)\equiv\mathcal{O}(\epsilon^{n})$, where $\epsilon$ is a small dimensionless parameter such that $U\sim\epsilon^{2}$, with $U$ minus the Newtonian potential. The metric is written in the harmonic gauge, $g^{\mu\nu}\Gamma^{\alpha}_{\mu\nu}=0$, under which the post-Newtonian EE take the form
\begin{equation}
    \left(-\partial_{t}^{2}+\nabla^{2}\right)w^{0}=-4\pi \sigma+\mathcal{O}(4), 
\end{equation}
\begin{equation}
    \nabla^{2}w^{i}=-4\pi \sigma^{i}+\mathcal{O}(2),
\end{equation}
where $\sigma=T^{00}+T^{jj}$, $\sigma^{i}=T^{0i}$ represent the effective mass and mass-current densities, respectively. 
Under the assumption of asymptotic flatness, the formal solutions are
\begin{equation}
    w^{0}(t,\mathbf{x})=\int \frac{\sigma(t,\mathbf{x}')}{\lvert\mathbf{x}-\mathbf{x'}\rvert}d^{3}x'+\frac{1}{2}\partial_{t}^{2}\int \sigma(t,\mathbf{x}')\lvert\mathbf{x}-\mathbf{x'}\rvert d^{3}x',
\end{equation}
\begin{equation}
    w^{i}(t,\mathbf{x})=\int \frac{\sigma^{i}(t,\mathbf{x}')}{\lvert\mathbf{x}-\mathbf{x'}\rvert}d^{3}x',
\end{equation}
which can equivalently be expressed in terms of the retarded time $t_\text{ret}=t-|\mathbf{x}-\mathbf{x}'|$. A great advantage of this approach is that the field equations are formally linear, allowing the application of the superposition principle to obtain the potentials generated by an $N$-body system: 
\begin{equation}
    w^{0}(t,\mathbf{x})=\sum_{A=1}^{N}w^{0}_{A}(t,\mathbf{x}), \qquad w^{i}(t,\mathbf{x})=\sum_{A=1}^{N}w^{i}_{A}(t,\mathbf{x}).
\end{equation}
The metric tensor does not fix the orientation of the spatial axes uniquely, but only up to some constant, time-independent, rotation about the origin. In the IAU realization, the axes are chosen to be kinematically non-rotating with respect to distant quasars and active galactic nuclei, which are assumed to have no proper motion, resulting in the International Celestial Reference System (ICRS). \\
The construction of the BCRS illustrates a crucial conceptual limitation of the post-Newtonian approach: the field equations are linear, and the metric potentials are treated as additive quantities superposed over a flat background. This hides the intrinsically nonlinear nature of the EE, where the gravitational field itself gravitates and interacts with matter and motion in a nontrivial way.
The construction of the BCRS highlights a conceptual limitation of the post-Newtonian approach. The nonlinear character of GR is incorporated only perturbatively: although quadratic and higher-order terms appear in the metric, the field equations governing the potentials are formally linear and therefore admit superposition at the level of $w^{0}$ and $w^{i}$.
The entire framework is built upon a fixed Minkowski background and a hierarchical expansion in powers of a small parameter, so that the fully nonlinear structure of the EE is recovered only order by order, rather than being treated exactly from the outset.
By contrast, in stationary axisymmetric solutions of the EE, such nonlinearity manifests through frame dragging and non-trivial asymptotic conical geometry that have no consistent Newtonian analogue. The BCRS, therefore, provides an accurate relativistic reference system only within the weak-field, quasi-isolated regime of the Solar System; its extension to galactic scales cannot be assumed. \\
A further obstacle to the use of the BCRS in the context of general relativistic galaxy models is that its definition presupposes a spacetime metric valid globally and the existence of objects with vanishing proper motion, which, within GR, can only be regarded as idealized, purely mathematical constructs. What is instead required is a genuinely local reference system applicable beyond the Solar System, covering the region of spacetime in which the general relativistic galactic models are assumed to be valid, and operationally realizable through astronomical observations, that can be defined without assumptions about motion outside the domain of validity of the model and without imposing any asymptotic structure on the spacetime.

%%%%%%%%%%%%%%%%%%%%%%%%%%%%%%%%%%%%%

\subsection{Zero angular momentum observers}
\label{ZAMO}

In the context of stationary, axisymmetric solutions of the EE, a commonly adopted reference system is that of Zero Angular Momentum Observers (ZAMOs). One of its advantages is that the existence and definition of ZAMOs rely only on the local geometry of the spacetime, independent of any asymptotic condition. One does not need to invoke distant stars or an inertial frame at infinity to define a ZAMO; the condition of zero angular momentum is purely local. This makes ZAMOs well-defined even in spacetimes that are not asymptotically flat or when focusing on a localized region of a galaxy.
These observers follow spatial circular orbits and have zero angular momentum, implying that their four-velocity is orthogonal to the hypersurfaces $t=\text{const.}$. The four-velocity of a ZAMO can be written as 
\begin{equation}
    u_Z=\frac{1}{\sqrt{-H}}(\partial_t+\chi\partial_\phi),
\end{equation}
where the angular velocity $\chi$ is fixed by the condition $L=0$, with $L$ denoting the conserved angular momentum~\eqref{conserved quantities} associated to the spacelike Killing vector $\psi=\partial_{\phi}$
\begin{equation}
    L=g_{\mu\nu}\psi^{\mu}u_{Z}^{\nu}=\frac{1}{\sqrt{-H}}(g_{t\phi}+\chi g_{\phi\phi}).
\end{equation}
Enforcing $L=0$ yields $\chi=-\frac{g_{t\phi}}{g_{\phi\phi}}$. In the case of the BG model this reads $\chi_{BG}=-\frac{\eta}{\eta^{2}-r^{2}}$. Notice that the ZAMOs cannot be defined globally, since they become spacelike in the region $\eta<r$. Indeed,
\begin{equation}
    g_{\alpha\beta}u_{Z}^{\alpha}u_{Z}^{\beta}=\frac{r^{2}}{g_{\phi\phi}H}=\frac{r^{2}}{r^{2}-\eta^{2}}>0 \,\, \text{if} \,\, \eta<r.
\end{equation}
This limitation, however, is not a problem since the domain $\eta<r$ is already excluded by the assumptions underlying the validity of the model, as discussed in Section~\ref{BG model}. \\
Considering a congruence of ZAMOs, the expansion vanishes: $\theta=\nabla_{\alpha}u^{\alpha}=0$. The shear is then given by
\begin{equation}
    \sigma_{\alpha\beta}=\nabla_{(\beta} u_{\alpha)}=\frac{1}{2}(\partial_{\alpha}u_{\beta}+\partial_{\beta}u_{\alpha}-2\Gamma^{\mu}_{\alpha\beta}u_{\mu}),
\end{equation}
whose non-vanishing components are
\begin{align}
    &\sigma_{tr}=\frac{1}{2}(e^{\nu}\nu,_{r}-e^{2\psi-\nu}\chi\chi,_{r}) \qquad \sigma_{tz}=\frac{1}{2}(e^{\nu}\nu,_{z}-e^{2\psi-\nu}\chi\chi,_{z}), \\
    &\sigma_{\phi r}=\frac{1}{2}e^{2\psi-\nu}\chi,_{r}, \qquad \sigma_{\phi z}=\frac{1}{2}e^{2\psi-\nu}\chi,_{z}.
\end{align}
Since the ZAMOs are hypersurface orthogonal, their vorticity vanishes. The non-zero components of the acceleration are
\begin{equation}
    a_{r}=\nu,_{r}, \qquad a_{z}=\nu,_{z}.
\end{equation}
In the case of rigid rotation the acceleration vanishes, while the shear is different from zero, as occurs also for the BG model. As a consequence, ZAMOs are not inertial observers. \\
An orthonormal tetrad associated to ZAMOs is
\begin{equation}
    e_{0}=\frac{\sqrt{g_{\phi\phi}}}{r}(\partial_{t}+\chi\partial_{\phi}), \quad
    e_{1}=e^{-\frac{\mu}{2}}\partial_{r}, \quad e_{2}=e^{-\frac{\mu}{2}}\partial_{z}, \quad e_{3}=-\frac{1}{\sqrt{g_{\phi\phi}}}\partial_{\phi}.
\end{equation}
Because the ZAMO congruence is well defined throughout the spacetime wherever it is timelike, at each such event, there exists a ZAMO observer against which the dust four-velocity can be computed. The relative velocity of the dust with respect to a ZAMO at the same event follows from~\eqref{relative velocity same event},
\begin{equation}
    v_{Z}=\frac{u}{\gamma_{Z}}-u_{Z}=\frac{\sqrt{g_{\phi\phi}}}{r}(\partial_{t}+(\Omega-\chi)\partial_{\phi}),
\end{equation}
where
\begin{equation}
    \gamma_{Z}=-g(e_{0},u)=\frac{r}{\sqrt{-H}\sqrt{g_{\phi\phi}}}.
\end{equation}
From here one can obtain the relative velocity of the dust with respect to a ZAMO at the same event using~\eqref{velocity gamma},
\begin{equation}
    v_{Z}=\frac{\eta}{r}.
\end{equation}
By evaluating this local quantity along circular dust orbits at different radii, one obtains a profile $v_{Z}(r)$, which is often interpreted as the general relativistic analogue of a galactic rotation curve. This has been tested on the Milky Way from \num{5} up to \SI{19}{\kilo\parsec} from the galactic center by fitting the data of a subsample made of \num{719 143} stars (including \num{241 918} OBA stars, \num{475 520} RGB giants, and \num{1705} Cepheides) selected from a sample of nearly $33$ million stars from the Gaia DR2 and DR3 catalogs~\cite{Crosta:2020,Beordo:2024}. The results show that the BG velocity profile is statistically indistinguishable from its state-of-the-art DM analogues. \\
This analysis relies on a rigorous formulation of the attitude reference system used to relate mathematical quantities to astronomical observables, accounting for all corrections due to the position and motion of the Gaia satellite with respect to the BCRS, which has been developed in e.g.~\cite{Bini:2003wv, Bini:2003ww, deFelice:2006xm, Crosta2019}.

%%%%%%%%%%%%%%%%%%%%%%%%%%%%%%%%%%%%%%%%%%%

\subsection{Locally inertial reference system}
\label{Locally inertial frame}

\begin{figure}
\centering
\includegraphics[scale=0.6]{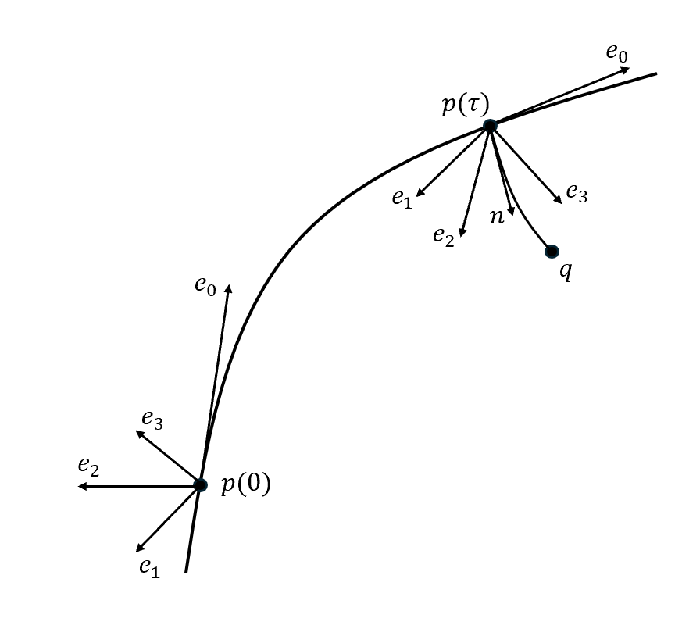}
\caption{Schematic illustration of the construction of the locally inertial reference system.}
\label{fig: Fermi coordinates}
\end{figure}

A locally inertial reference system translates the Newtonian notion of an inertial framework to the neighborhood of an observer’s worldline, providing the most appropriate setting in which to describe physical quantities. Because all measurements are performed in a small neighborhood of the observer, one introduces a rest frame adapted to that motion, consisting of a clock measuring proper time and a triad of orthonormal spatial axes. The orientation of this triad is, in principle, arbitrary, but can be set according to the physical situation under study. A natural choice is to require the system to remain locally inertial, so that no fictitious forces appear in the equations of motion. This condition is ensured by Fermi–Walker transport of the spatial triad along the observer’s trajectory, which can be operationally realized by means of three mutually orthogonal gyroscopes. \\
This approach contrasts with the construction of the BCRS, whose spatial axes are fixed by assuming an asymptotically flat spacetime and by identifying distant astronomical sources with negligible proper motion. Such axes are kinematically non-rotating with respect to those distant objects, but in general rotate relative to gyroscopes carried by a local observer. Consequently, the BCRS is inertial only asymptotically, and motion described within it necessarily includes fictitious forces. \\
In this work we instead construct a dynamically non-rotating reference system: its spatial axes follow the torque-free directions defined by gyroscopes, ensuring a genuinely local inertial frame. This yields a laboratory reference system that can be operationally defined within the spacetime region of interest, without relying on assumptions about the behavior of the spacetime outside that domain. \\
The procedure to construct the locally inertial reference system can be summarized as follows~\cite{Manasse:1963zz, Gravitation}:
\begin{enumerate}
    \item Consider $x^{\alpha}$, $\alpha\in\{0,1,2,3\}$, a generic coordinate system in terms of which the expression of the metric is known. Let $x^{\alpha}(\tau)$ describe the world line of an observer parameterized by the proper time $\tau$ and let its timelike tangent unit vector $u^{\alpha}=\frac{d x^{\alpha}}{d \tau}$ be the observer's four-velocity (see Section~\ref{geodesics}). 
    \item Build the tetrad adapted to the observer. 
    We choose a local basis such that at the point $p(\tau)$ occupied by the dust particle the metric is the flat one; the timelike vector is the vector tangent to the observer's trajectory $e_{0}=u$, while the spatial triad $e_{i}$, $i={1,2,3}$ (orthogonal to $u$) is obtained by Fermi-Walker transport along the observer's worldline:
    \begin{align}
    \frac{de_{i}^{\alpha}}{d\tau}=-(a^{\alpha}u_{\beta}-u^{\alpha}a_{\beta})e_{i}^{\beta},
    \end{align}
    where $a^{\alpha}=\frac{du^{\alpha}}{d\tau}$ is the observer's acceleration. For an observer following a geodesic, the acceleration is zero, and the above equation reduces to parallel transport. Since the metric has vanishing Fermi-Walker derivative along any curve, in addition to preserving the orthogonal decomposition of the tangent space parallel and perpendicular to $u$ along the curve, all inner products are invariant under this transport. We are interested in Fermi-Walker transport because it describes the behavior of the spin vector of a torque-free gyroscope carried by an observer along its world line. Therefore, the spatial vectors $e_{i}$ of a Fermi frame along the world line are locally non-rotating with respect to a set of three independently oriented gyroscopes carried by the observer and span the associated local rest space (see Section~\ref{gyroscopes}). 
    \item Consider now all spacelike geodesics which pass through the generic point $p(\tau)$ on the observer's world line and which are orthogonal to $u$ at $p(\tau)$. These spacelike geodesics form a hypersurface in a normal neighborhood of the observer's world line. \\
    Let $q$ with coordinates $x^{\alpha}$ be a generic spacetime point in the vicinity of $p$ on such a hypersurface and consider the unique spacelike geodesic segment from $p$ to $q$ of proper length $s$. The Fermi coordinates $(T,X^{1},X^{2},X^{3})$ of $q$ are then given by 
    \begin{equation}
        T=\tau, \quad X^{i}=sg_{\alpha\beta}n^{\alpha}e_{i}^{\beta}|_{q},
    \end{equation}
    where $n$ is the unit tangent vector to the spacelike geodesic segment at $p(\tau)$ satisfying the condition $g_{\alpha\beta}n^{\alpha}u^{\beta}|_{q}=0$ (see Section~\ref{Fermi coordinates}). A schematic representation of the construction is shown in Figure~\ref{fig: Fermi coordinates}.
    \item In the points around $p(\tau)$, up to distances $\lambda$ such that $|R_{\mu\nu\rho\sigma}|\lambda^2\ll1$\footnote{With $|\cdot|$ for a tensor, we intend the maximum of the modulus among the components of the tensor at every point.}, the metric takes the form
    \begin{align}
    \label{locally inertial metric}
    ds^2=-\left(1+R_{TjTk}X^jX^k\right)dT^2-\frac 43 R_{Tjlk}X^jX^k\ dTdX^l+\left(\delta_{mn}-\frac 13R_{mjnk}X^jX^k\right) dX^m dX^n,
    \end{align}
    where $R_{abcd}$ is the Riemann tensor in the Fermi frame. Eq.~\eqref{locally inertial metric} gives the metric in a neighborhood of an observer up to second order in the Fermi coordinates (see Section \ref{metric Fermi}).
\end{enumerate}

The applicability of Fermi coordinates is restricted by the local curvature of spacetime. When comparing measurements performed by observers at widely separated locations within the Galaxy, or even in different galaxies, different local coordinate patches must be matched consistently. This requires the corresponding observers to adopt a common convention for the orientation of their spatial axes. In Section~\ref{radially locked reference system} we propose to achieve this by locking the triad of locally inertial observers to the radial direction defined by photons coming from the galactic center. This construction defines what we call a radially locked tetrad. Physically, the radial direction should be identified with the spatial component of the four-momentum of photons passing through the center with zero angular momentum. Due to frame dragging, this direction does not coincide with $\partial_{r}$. From an operational standpoint, it may therefore be more convenient to rotate the spatial triad of the locally inertial frame of a geodetic observer so that one axis always points toward the galactic center as represented in its own Fermi coordinates.

%%%%%%%%%%%%%%%%%%%%%%%%%%%%%%%%%

\subsubsection{Geodetic motion in the galactic plane}
\label{geodesics}

The construction of a locally inertial reference system begins with the identification of the observer's worldline. In our case, the observer is attached to a dust particle moving along a geodesic confined to the galactic plane $z=0$. We therefore start by examining the geodesic equations associated with the metric~\eqref{ZAMO metric}:
\begin{subequations}
\begin{align}
        &\Ddot{t}+2\left(\nu,_{r}-\frac{1}{2}e^{2(\psi-\nu)}\chi\chi,_{r}\right)\dot{t}\dot{r}+2\left(\nu,_{z}-\frac{1}{2}e^{2(\psi-\nu)}\chi\chi,_{z}\right)\dot{t}\dot{z}+e^{2(\psi-\nu)}\chi,_{r}\dot{\phi}\dot{r}+e^{2(\psi-\nu)}\chi,_{z}\dot{\phi}\dot{z}=0, \\
        &\ddot{\phi}+2\left(-\chi\psi,_{r}-\frac{1}{2}\chi,_{r}+\chi\nu,_{r}-\frac{1}{2}e^{2(\psi-\nu)}\chi^{2}\chi,_{r}\right)\dot{t}\dot{r}+2\left(-\chi\psi,_{z}-\frac{1}{2}\chi,_{z}+\chi\nu,_{z}-\frac{1}{2}e^{2(\psi-\nu)}\chi^{2}\chi,_{z}\right)\dot{t}\dot{z}+ \notag \\
        &+2\left(\psi,_{r}+\frac{1}{2}e^{2(\psi-\nu)}\chi\chi,_{r}\right)\dot{\phi}\dot{r}+2\left(\psi,_{z}+\frac{1}{2}e^{2(\psi-\nu)}\chi\chi,_{z}\right)\dot{\phi}\dot{z}=0, \\
        &e^{\mu}\ddot{r}+\nu,_{r}e^{2\nu}\dot{t}^{2}-\psi,_{r}e^{2\psi}(\dot{\phi}-\chi\dot{t})^{2}+\chi,_{r}e^{2\psi}(\dot{\phi}-\chi\dot{t})\dot{t}+\frac{1}{2}\mu,_{r}e^{\mu}(\dot{r}^{2}-\dot{z}^{2})+\mu,_{z}e^{\mu}\dot{r}\dot{z}=0,
        \label{eq: geodesic r}\\
        &e^{\mu}\ddot{z}+\nu,_{z}e^{2\nu}\dot{t}^{2}-\psi,_{z}e^{2\psi}(\dot{\phi}-\chi\dot{t})^{2}+\chi,_{z}e^{2\psi}(\dot{\phi}-\chi\dot{t})\dot{t}-\frac{1}{2}\mu,_{z}e^{\mu}(\dot{r}^{2}-\dot{z}^{2})+\mu,_{r}e^{\mu}\dot{r}\dot{z}=0
        \label{eq: geodesic z},
    \end{align}
\end{subequations}
where an overdot indicates differentiation with respect to an affine parameter. As discussed in Section \ref{Stationary axisymmetric spacetimes}, stationarity and axial symmetry imply that $t$ and $\phi$ are cyclic variables, giving two constants of motion
\begin{align}
 E&=e^{2\nu} \dot t +e^{2\psi} (\dot\phi-\chi\dot t)\chi, \\
 L&=e^{2\psi} (\dot\phi-\chi\dot t).
\end{align}
Eliminating $\dot{\phi}$ one finds $\dot t=e^{-2\nu} (E-L\chi)$. Moreover, we can choose the affine parameter such that
\begin{equation}
    g_{\mu\nu}\frac{dx^{\mu}}{d\lambda}\frac{dx^{\nu}}{d\lambda}=\epsilon=
    \begin{cases}
        -1 \,\,\, \text{timelike geodesic} \\
        \,\,\,\,\,0 \,\,\, \text{null geodesic} \\
        +1 \,\,\, \text{spacelike geodesic}
    \end{cases}
\end{equation}
which explicitly reads
\begin{equation}
\label{constraint}
        -e^{-2\nu}(E-\chi L)^{2}+e^{-2\psi}L^{2}+e^{\mu}(\dot{r}^{2}+\dot{z}^{2})=\epsilon.
    \end{equation}
This relation can be interpreted as a generalized ``energy balance'' equation, where the first two terms play the role of an effective potential containing both gravitoelectric and gravitomagnetic contributions, while the last term represents the kinetic energy in the meridional plane~\cite{Astesiano:2023emg}. It serves as a first integral of the last two geodesic equations, so that the system reduces to
\begin{subequations}
\begin{align}
        \label{integrated geodesic eq t}
        &\dot{t}=e^{-2\nu}(E-\chi L), \\
        \label{integrated geodesic eq phi}
        &\dot{\phi}=\chi e^{-2\nu}(E-\chi L)+Le^{-2\psi}, \\
        \label{integrated geodesic eq r,z}
        &-e^{-2\nu}(E-\chi L)^{2}+e^{-2\psi}L^{2}+e^{\mu}(\dot{r}^{2}+\dot{z}^{2})=\epsilon,
\end{align}
\end{subequations}
together with either~\eqref{eq: geodesic r} or~\eqref{eq: geodesic z}. \\
We now focus on planar motions characterized by $\dot{z}=0$, and in particular on the timelike geodesics ($\epsilon=-1$) since these describe the orbits of dust particles forming the disc, and to which local observers can be attached. In stationary, axisymmetric dust configurations such trajectories are indeed admissible, as the symmetry guarantees confinement to the planes $z=\text{const.}$ From conservation of the stress-energy tensor ($\nabla^{\mu}T_{\mu\nu}=0$) we obtain
\begin{equation}
    \nabla^{\mu}(\rho u_{\mu})=0, \quad u_{\mu}\nabla^{\mu}u_{\nu}=0.
\end{equation}
The circularity condition~\eqref{circularity condition four-velocity} forces the four-velocity of the dust to be of the form~\eqref{dust velocity}, corresponding to circular trajectories at constant $z$. We can therefore conclude that dust particles follow geodesics restricted to such planes. \\
By setting $z=\text{const.}$ in the $r$ and $z$ components of the geodesic equations we obtain 
\begin{equation}
 \nu,_z e^{-2\nu} (E-L\chi)^2-\psi,_z e^{-2\psi} L^2+L\chi,_z e^{-2\nu} (E-L\chi)-\frac 12 \mu,_z e^\mu \dot r^2 = 0.
\end{equation}
This equation, together with the constraint~\eqref{constraint} and the conserved quantities specify the planar motion. As a consistency check, the geodesics of dust particles are recovered as special solutions specified by
\begin{align}
\label{dust geodesic}
 \dot t=\frac 1{\sqrt {-H}}, \qquad \dot \phi=\frac \Omega{\sqrt {-H}}, \qquad r=r_0, \qquad z=z_0.
\end{align}
with corresponding constants of motion
\begin{align}
  \left. L=\frac {\eta}{\sqrt {-H}}\right|_{r=r_0, z=z_0}, \qquad \left.E=\Omega L+\sqrt{-H}\right|_{r=r_0, z=z_0}.
\end{align}

%%%%%%%%%%%%%%%%%%%%%%%%%%%%%%%%%%%%%%%%

\subsubsection{Gyroscopes}
\label{gyroscopes}

After specifying the observer's worldline, the next step in constructing an adapted locally inertial reference system is to introduce an orthonormal tetrad, with the spatial vectors aligned with the directions of orthogonal gyroscopes. For a torque-free gyroscope carried along a geodesic, Fermi-Walker transport reduces to the parallel transport of its spin four-vector $S^{\mu}$~\cite{MisnerThorneWheeler1973}:
\begin{equation}
    \dot{S}^{\mu}+\Gamma^{\mu}_{\nu\rho}S^{\nu}\dot{x}^{\rho}=0.
\end{equation}
After some algebra, this equation can be written in a more explicit form, adapted to a dust geodesic~\eqref{dust geodesic}:
\begin{subequations}
\begin{align}
    &\dot{S}^{t}+\left[\left(1-\frac{\eta^{2}}{r^{2}}\right)\nu,_{r}+\frac{\eta^{2}}{r^{3}}\right]\frac{S^{r}}{2\sqrt{-H}}+\left(1-\frac{\eta^{2}}{r^{2}}\right)\nu,_{z}\frac{S^{z}}{2\sqrt{-H}}=0, \\
    &\dot{S}^{\phi}+\left(1-\frac{\eta^{2}}{r^{2}}\right)\left(\Omega-\frac{H}{\eta}\right)\nu,_{z}\frac{S_{z}}{2\sqrt{-H}}+\left[\left(1-\frac{\eta^{2}}{r^{2}}\right)\nu,_{r}+\frac{\eta^{2}}{r^{3}}\right]\left(\Omega-\frac{H}{\eta}\right)\frac{S^{r}}{2\sqrt{-H}}=0, \\
    &\dot{S}^{r}+\frac{e^{-\mu}}{2\sqrt{-H}}\left(\frac{r^{2}-\eta^{2}}{\eta}\nu,_{r}+\frac{\eta}{r}\right)(\Omega S^{t}-S^{\phi})=0, \\
    &\dot{S}^{z}+\frac{e^{-\mu}}{\sqrt{-H}}\frac{r^{2}-\eta^{2}}{\eta}\frac{\nu,_{z}}{2}(\Omega S^{t}-S^{\phi})=0.
\end{align}
\end{subequations}
For a dust geodesic as above, all coefficients in this system are evaluated at $r=r_0$, $z=z_0$, and are therefore constant. The gyroscope equations can thus be solved straightforwardly. It is interesting to notice that $\dot S^z$ is different from zero in general (also for the BG model), leading to precession in the $z-\phi$ plane.
This is expected, since the $z$-gradient of the velocity of the particles is non-zero in general. However, if we restrict to motions in the equatorial plane ($z=0$) and assume symmetry with respect to it, then at $z=0$ we have $\nu,_z=0$ and the equation for $S^z$ reduces to $\dot S^z=0$. In what follows, we shall impose this restriction only later, and for now, solve the gyroscope equations for an arbitrary circular motion of a dust particle. \\
After introducing the vector
\begin{align}
 S=
\begin{pmatrix}
 S^t \\ S^z \\ S^r \\ S^\phi
\end{pmatrix},
\end{align}
we can write the equations in the form
\begin{equation}
\label{gyroscope equations matrix form}
 \dot S=MS,
\end{equation}
where $M$ is the constant (along the circular orbits) matrix~\eqref{M matrix}.
The solution of Eq.~\eqref{gyroscope equations matrix form} is then
\begin{align}
 S(\tau)=e^{M \tau} S_0,
\end{align}
where 
\begin{align}
 S_0=\begin{pmatrix}
 S^t_0 \\ S^z_0 \\ S^r_0 \\ S^\phi_0
\end{pmatrix},
\end{align}
is the initial configuration (at $\tau=0$), and 
\begin{align}
  e^{M\tau} =
\begin{pmatrix}
 -\frac {x\cos (\omega\tau) }{{yz-x}} +\frac {yz}{yz-x} & \frac {\alpha \sin (\omega\tau)}{\sqrt {\alpha^2+\beta^2}\sqrt{yz-x}} & \frac {\beta \sin (\omega\tau)}{\sqrt {\alpha^2+\beta^2}\sqrt{yz-x}} & -\frac {y(1-\cos(\omega\tau))}{yz-x} \\
 \frac {\alpha x \sin (\omega\tau)}{\sqrt {\alpha^2+\beta^2}\sqrt{yz-x}} & \frac {\beta^2}{\alpha^2+\beta^2}+\frac {\alpha^2 \cos (\omega\tau)}{\alpha^2+\beta^2} & -\frac {\alpha \beta (1-\cos(\omega\tau))}{\alpha^2+\beta^2} & -\frac {\alpha y \sin (\omega\tau)}{\sqrt {\alpha^2+\beta^2}\sqrt{yz-x}} \\
 \frac {\beta x \sin (\omega\tau)}{\sqrt {\alpha^2+\beta^2}\sqrt{yz-x}} &  -\frac {\alpha \beta (1-\cos(\omega\tau))}{\alpha^2+\beta^2} & \frac {\alpha^2}{\alpha^2+\beta^2}+\frac {\beta^2 \cos (\omega\tau)}{\alpha^2+\beta^2} & -\frac {\beta y \sin (\omega\tau)}{\sqrt {\alpha^2+\beta^2}\sqrt{yz-x}}\\
 \frac {xz(1-\cos(\omega\tau))}{ {yz-x}} & \frac {\alpha z \sin (\omega\tau)}{\sqrt {\alpha^2+\beta^2}\sqrt{yz-x}} & \frac {\beta z \sin (\omega\tau)}{\sqrt {\alpha^2+\beta^2}\sqrt{yz-x}} & -\frac x{yz-x}+\frac {yz\cos (\omega\tau)}{yz-x}
\end{pmatrix}.
\label{solution gyroscope equation}
\end{align}
We refer the reader to Appendix~\ref{A: Gyroscope equation solution} for the details of the derivation and the definitions of the quantities appearing in the above expression. \\
This result is completely general and applies to the motion of gyroscopes attached to any dust particle. However, as remarked in Section~\ref{geodesics}, we restrict to motions on the equatorial plane, assuming reflection symmetry with respect to it. In this case, all $z$ derivatives vanish, and the exponential matrix simplifies to
\begin{align}
  e^{M\tau}|_{z=0} =&
\begin{pmatrix}
 \frac {\Omega\eta(\cos (\omega\tau)-1)}{H} +1 & 0 & \frac {\eta e^{\frac \mu2}\sin (\omega\tau)}{r\sqrt{-H}} & \frac {\eta(1-\cos(\omega\tau))}{H} \\
 0 & 1 & 0 & 0\\
 \frac {\Omega r e^{-\frac \mu2} \sin (\omega\tau)}{\sqrt{-H}} &  0 & \cos (\omega\tau) & -\frac {r e^{-\frac \mu2} \sin (\omega\tau)}{\sqrt{-H}}\\
 -\frac {\Omega (\Omega\eta-H)(1-\cos(\omega\tau))}{{H}} & 0 & \frac {(\Omega\eta-H)e^{\frac \mu2} \sin (\omega\tau)}{r\sqrt{-H}} & \frac {\Omega\eta (1-\cos (\omega\tau))}H +\cos (\omega\tau)
\end{pmatrix}.
\label{solution gyroscope equations galactic plane}
\end{align}
This analysis provides the exact law governing the evolution of gyroscopes transported along the geodesics of the dust congruence. The exponential matrix $e^{M\tau}$ encodes the precession of the spin vector relative to the coordinate basis, with the frequency $\omega$ characterizing the rate at which local inertial directions rotate with respect to the axis of a static observer due to spacetime curvature and rotation. In the equatorial plane, this precession reduces to a rotation in the $r-\phi$ plane. Operationally, this means that a gyroscope carried along a dust worldline rotates with constant angular velocity relative to the symmetry axis. If one wished to keep the gyroscope aligned toward that axis, it would have to be continuously counter–rotated by the same angular velocity. An observer attached to the dust particle would therefore infer that their local inertial frame is steadily rotating about the axis of symmetry. Physically, this effect represents the general relativistic frame dragging produced by the galaxy’s rotating mass distribution, namely the fully relativistic generalization of the Lense–Thirring precession that arises in the weak–field, slow–motion limit~\cite{Lense:1918zz,1995grin.book.....C,Iorio2010}. Hence, the precession of gyroscopes along dust worldlines provides a direct probe of the non–Newtonian structure of the spacetime, revealing how local inertial frames are twisted by the global rotation.

%%%%%%%%%%%%%%%%%%%%%%%%%%%%%%%%%%

\subsubsection{Tetrad adapted to locally inertial observers}
\label{locally inertial tetrad}

We now have all the elements to construct the tetrad adapted to a locally inertial observer comoving with a dust particle in the galactic plane. This tetrad defines the locally inertial reference system carried by a geodesic observer, in which all kinematical quantities vanish identically. This is in contrast with the BCRS and the ZAMO reference systems, which are only asymptotically inertial: their kinematical quantities vanish far from the source but not in its interior, where their axes rotate with respect to gyroscopes due to frame dragging. \\
We start by choosing a point on the dust geodesic corresponding to $\tau=0$ and imposing the metric there to be the Minkowski one. Then set the timelike vector of the tetrad to be the tangent to the geodesic at that point,
\begin{equation}
    e_{0}=u=\frac{1}{\sqrt{-H}}(\partial_{t}+\Omega\partial_{\phi}).
\end{equation}
The remaining three vectors defining the reference system follow from the orthonormality condition, $g_{\mu\nu}|_{G}=g_{\alpha'\beta'}e^{\alpha'}_{\mu}e^{\beta'}_{\nu}=\eta_{\mu\nu}$, giving
\begin{subequations}
\begin{align}
    e_{1}&=e^{-\frac{\mu}{2}}\partial_{z}, \\
    e_{2}&=e^{-\frac{\mu}{2}}\partial_{r}, \\
    e_{3}&=\frac{1}{\sqrt{-H}}(\frac{\eta}{r}\partial_{t}+\frac{\eta\Omega-H}{r}\partial_{\phi}).
\end{align}
\end{subequations}
As described previously, the spatial triad at any proper time $\tau$ is determined by the solution of the gyroscope equations $S(\tau)=e^{M\tau}S_{0}$, with $e^{M\tau}$ restricted to the galactic plane as in equation~\eqref{solution gyroscope equations galactic plane} and $S^{0}$ given by the initial tetrad above. The result is 
\begin{subequations}
\begin{align}
\label{eq: locally inertial tetrad}
    e_{0}&=\frac{1}{\sqrt{-H}}(\partial_{t}+\Omega\partial_{\phi}), 
    \\
    e_{1}&=e^{-\frac{\mu}{2}}\partial_{z}, 
    \\
    e_{2}(\tau)&=\frac{\eta}{r\sqrt{-H}}\sin{(\omega\tau)}\partial_{t}+e^{-\frac{\mu}{2}}\cos{(\omega\tau)}\partial_{r}+\frac{(\eta\Omega-H)}{r\sqrt{-H}}\sin{(\omega\tau)}\partial_{\phi}, 
    \\
    e_{3}(\tau)&=\frac{\eta}{r\sqrt{-H}}\cos{(\omega\tau)}\partial_{t}-e^{-\frac{\mu}{2}}\sin{(\omega\tau)}\partial_{r}+\frac{(\eta\Omega-H)}{r\sqrt{-H}}\cos{(\omega\tau)}\partial_{\phi}.
\end{align}
\end{subequations}
The spatial axes $e_{2}$ and $e_{3}$ undergo precession in the galactic plane with angular frequency $\omega$, representing the rotation of the inertial directions of gyroscopes with respect to kinematically non-rotating spatial axes. \\
For completeness, the coordinate basis can be expressed in terms of the locally inertial tetrad as
\begin{subequations}
\begin{align}
    \label{change of basis t}
    \partial_{t}&=\frac{\Omega\eta-H}{\sqrt{-H}}e_{0}-\frac{\Omega r}{\sqrt{-H}}(\sin{(\omega\tau)}e_{2}+\cos{(\omega\tau)}e_{3}), \\
    \label{change of basis phi}
    \partial_{\phi}&=\frac{-\eta}{\sqrt{-H}}e_{0}+\frac{r}{\sqrt{-H}}(\sin{(\omega\tau)}e_{2}+\cos{(\omega\tau)}e_{3}), \\
    \label{change of basis r}
    \partial_{r}&=e^{\mu/2}(\cos{(\omega\tau)}e_{2}-\sin{(\omega\tau)}e_{3}).
\end{align}
\end{subequations}
The particular case of rigidly rotating dust, including the BG model, is simply obtained by setting $\Omega=0$, $H=-1$ in the above expressions, yielding
\begin{subequations}
\begin{align}
    e_{0}&=\partial_{t}, \\
    e_{1}&=e^{-\frac{\mu}{2}}\partial_{z}, \\
    e_{2}&=\frac{\eta}{r}\sin{(\omega\tau)}\partial_{t}+e^{-\frac{\mu}{2}}\cos{(\omega\tau)}\partial_{r}+\frac{1}{r}\sin{(\omega\tau)}\partial_{\phi}, \\
    e_{3}&=\frac{\eta}{r}\cos{(\omega\tau)}\partial_{t}-e^{-\frac{\mu}{2}}\sin{(\omega\tau)}\partial_{r}+\frac{1}{r}\cos{(\omega\tau)}\partial_{\phi}.
\end{align}
\end{subequations}
The precession of $e_{2}$ and $e_{3}$ with respect to the fixed axes of the BCRS reveals that such a reference system is not inertial in the interior of the solution. Although its spatial axes are anchored to distant non-rotating objects, they rotate locally with respect to gyroscopes as a consequence of frame dragging.

%%%%%%%%%%%%%%%%%%%%%%%%%%%%%%%%%%%

\subsubsection{Fermi coordinates in the galactic plane}
\label{Fermi coordinates}

We now construct the coordinate system naturally associated with a locally inertial observer. Unlike the cylindrical coordinates used to describe the global geometry, Fermi coordinates provide a local description of spacetime from the viewpoint of a locally inertial observer (here, a dust particle moving on a circular spatial trajectory in the galactic plane) with spatial axes aligned with mutually orthogonal gyroscopes. In these coordinates, the metric reduces locally to its Minkowskian form, and curvature effects appear only at second order in the spatial distance from the observer's worldline. This coordinate system thus represents the most faithful general relativistic analogue of the Cartesian coordinates of an inertial observer in Newtonian mechanics~\cite{Manasse:1963zz,MisnerThorneWheeler1973}. \\
As discussed above, the time coordinate is the proper time of the observer along its worldline. For a dust particle geodesic~\eqref{dust geodesic}, we have 
$\frac{dt}{d\tau}=\frac{1}{\sqrt{-H}}$, so that
\begin{equation}
    T=\tau=\sqrt{-H_{0}}t,
\end{equation}
where $H_{0}:=H(r_{0},z_{0}=0)$. The spatial coordinates are defined by projecting the tangent vector of the unique spacelike geodesic, orthogonal to the observer's worldline and connecting the observer to the event whose coordinates are to be assigned, onto the observer's triad, $X^{i}=sg_{\mu\nu}n^{\mu}e_{i}^{\nu}|_{q}$, see figure~\ref{fig: Fermi coordinates}. Here $s$ is the proper distance along the spacelike geodesic and $n$ its tangent vector. To obtain $n$ we integrate the equations for spacelike geodesics, corresponding to the $\epsilon=1$ case in~\eqref{integrated geodesic eq t},~\eqref{integrated geodesic eq phi},~\eqref{integrated geodesic eq r,z}, which directly yield the components of the tangent vector to an arbitrary spacelike geodesic in the galactic plane:
\begin{subequations}
\begin{align}
    \frac{dt}{ds}&=e^{-2\nu}(E-\chi L), \label{eq t}\\
    \frac{dz}{ds}&=0, \label{eq z}\\
    \frac{dr}{ds}&=\pm e^{-\frac{\mu}{2}}[e^{-2\nu}(E-\chi L)^{2}-e^{-2\psi}L^{2}+1]^{\frac{1}{2}}, \label{eq r}\\
    \frac{d\phi}{ds}&=e^{-2\psi}L+\chi e^{-2\nu}(E-\chi L). \label{eq phi}
\end{align}
\end{subequations}
Remember that we only care about spacelike geodesics orthogonal to the observer's worldline, namely those whose tangent vector satisfies $g_{\mu\nu}n^{\mu}u^{\nu}=0$, corresponding to 
 \begin{equation}
    \dot{t}(g_{tt}+\Omega g_{t\phi})+\dot{\phi}(g_{t\phi}+\Omega g_{\phi\phi})=0,
    \end{equation}
which implies
\begin{equation}
    \frac{\dot{t}}{\dot{\phi}}=\frac{e^{2\psi}(\Omega-\chi)}{e^{2\nu}+\chi e^{2\psi}(\Omega-\chi)}.
\end{equation}
On the other hand the geodesic equations~\eqref{integrated geodesic eq t},~\eqref{integrated geodesic eq phi},~\eqref{integrated geodesic eq r,z} give
\begin{equation}
    \frac{\dot{t}}{\dot{\phi}}=\frac{e^{-2\nu}(E-\chi L)}{\chi e^{-2\nu}(E-\chi L)+Le^{-2\psi}}.
\end{equation}
Equating the two expressions and substituting the metric components in terms of $\eta$ and $H$ yields the following relation between $E$ and $L$ in terms of the metric functions,
\begin{equation}
\label{orthogonality constraint spacelike geodesics}
    E=\Omega L,
\end{equation}
which defines the spacelike geodesics orthogonal to the observer's worldline. The tangent vector in the coordinate basis is therefore
\begin{align}
    n=\frac{\eta}{r^{2}}L\partial_{t}+\frac{\Omega\eta-H}{r^{2}}L\partial_{\phi}\pm e^{-\frac{\mu}{2}}\sqrt{\frac{H}{r^{2}}L^{2}+1}\partial_{r}.
\end{align}
The Fermi spatial coordinates are then given by 
\begin{align}
    X&=sg_{\alpha\beta}n^{\alpha}e_{2}^{\beta}|_{q}=s\left[\frac{\sqrt{-H_{0}}L}{r_{0}}\sin{(\omega_{0}\tau)}+\sqrt{1+\frac{H_{0}L^{2}}{r^{2}}}\cos{(\omega_{0}\tau)}\right],
\end{align}
\begin{align}
    Y&=sg_{\alpha\beta}n^{\alpha}e_{3}^{\beta}|_{q}=s\left[\frac{\sqrt{-H_{0}}L}{r_{0}}\cos{(\omega_{0}\tau)}-\sqrt{1+\frac{H_{0}L^{2}}{r^{2}}}\sin{(\omega_{0}\tau)}\right],
\end{align}
\begin{equation}
\label{X1}
    Z=sg_{\alpha\beta}n^{\alpha}e_{1}^{\beta}|_{q}=0,
\end{equation}
where the subscript indicates quantities evaluated at the point $q$. \\
Instead of using cartesian Fermi coordinates $(X,Y,Z)$ it is often convenient to introduce local polar coordinates $(s,\theta)$, where $s$ is the proper distance
\begin{equation}
    s=\sqrt{X^{2}+Y^{2}},
\end{equation}
while $\theta$ is the angle between $n$ and the local $X-$axis $e_{2}$,
\begin{equation}
\label{n polar coordinates}
    n=\cos{\theta}e_{2}+\sin{\theta}e_{3}.
\end{equation}
Both $s$ and $\theta$ depend on the conserved energy and angular momentum along the spacelike geodesic, which are related by~\eqref{orthogonality constraint spacelike geodesics}. Expressing the tangent vector $n$ in the local basis $\{e_{0},e_{2},e_{3}\}$ on the galactic plane via equations~\eqref{change of basis t},~\eqref{change of basis phi},~\eqref{change of basis r} gives
\begin{align}
\label{n local reference system}
    n=&\left(\pm\sqrt{\frac{H}{r^{2}}L^{2}+1}\cos{(\omega\tau)}+\frac{\sqrt{-H}}{r}L\sin{(\omega\tau)}\right)e_{2} \notag \\
    &+\left(\mp\sqrt{\frac{H}{r^{2}}L^{2}+1}\sin{(\omega\tau)}+\frac{\sqrt{-H}}{r}L\cos{(\omega\tau)}\right)e_{3}.
\end{align}
Comparing with~\eqref{n polar coordinates}, we then obtain
\begin{align}
    &\cos{\theta}=\pm\sqrt{\frac{H}{r^{2}}L^{2}+1}\cos{(\omega\tau)}+\frac{\sqrt{-H}}{r}L\sin{(\omega\tau)}, \\
    &\sin{\theta}=\mp\sqrt{\frac{H}{r^{2}}L^{2}+1}\sin{(\omega\tau)}+\frac{\sqrt{-H}}{r}L\cos{(\omega\tau)},
\end{align}
so that 
\begin{equation}
    \theta=\arctan{\frac{\mp\sqrt{\frac{H}{r^{2}}L^{2}+1}\sin{(\omega\tau)}+\frac{\sqrt{-H}}{r}L\cos{(\omega\tau)}}{\pm\sqrt{\frac{H}{r^{2}}L^{2}+1}\cos{(\omega\tau)}+\frac{\sqrt{-H}}{r}L\sin{(\omega\tau)}}}.
\end{equation}
Inverting this relation shows that $\theta$ and $L$ are not in one-to-one correspondence. Indeed,
\begin{equation}
    L=\pm\frac{r}{\sqrt{-H}}\sin{(\theta+\omega\tau)}.
\end{equation}
Thus, for a given value of $L$ there correspond two angles $\theta$ separated by $\pi$. This reflects the fact that points on opposite sides of the observer's worldline are connected by the same spacelike geodesic. The equivalence of $n$ and $-n$ as tangent vectors, evident in~\eqref{n local reference system}, implies that this ambiguity in labeling symmetric points is also inherited by the local angular coordinate $\theta$.

%%%%%%%%%%%%%%%%%%%%%%%%%%%%%%%%%%%

\subsubsection{Metric tensor in Fermi coordinates}
\label{metric Fermi}

We can now write the metric in a neighborhood of the observer's geodesic as an expansion in Fermi coordinates. By construction, the metric reduces to the Minkowski form on the observer's geodesic, while its first derivatives vanish, ensuring that no inertial forces appear at the observer's position. The leading corrections therefore arise at quadratic order in the spatial Fermi coordinates and are entirely determined by the components of the Riemann tensor evaluated along the geodesic. \\
To write the metric up to second order in the Fermi coordinates, as in Eq.~\eqref{locally inertial metric}, we need first to write the Riemann tensor in the locally inertial reference system,
\begin{equation}
    R_{abcd}=R_{\alpha\beta\mu\nu}e^{\alpha}_{a}e^{\beta}_{b}e^{\mu}_{c}e^{\nu}_{d},
\end{equation}
with $\{a,b,c,d\}\in\{T,X,Y,Z\}$ and $\{\alpha, \beta, \mu, \nu\}\in\{t, z, r, \phi\}$. The explicit expression of the components of the Riemann tensor is reported in Appendix~\ref{A: Riemann locally inertial reference system}. The calculation is restricted to the galactic plane by setting $Z=0$ (see Eq.~\eqref{X1}) and imposing reflection symmetry with respect to it, resulting in $dZ=0$. The corresponding metric is
\begin{align}
\label{Fermi metric}
    ds^{2}|_{Z=0}=&-[1+R_{TXTX}X^{2}+2R_{TXTY}XY+R_{TYTY}Y^{2}]dT^{2} \notag \\
    &-\frac{4}{3}[R_{TXXY}XY+R_{TYXY}Y^{2}]dTdX \notag\\
    &-\frac{4}{3}[R_{TXXY}X^{2}+R_{TYXY}XY]dTdY \notag \\
    &+\left[1-\frac{1}{3}R_{XYXY}Y^{2}\right]dX^{2}+\left[1-\frac{1}{3}R_{XYXY}X^{2}\right]dY^{2} \notag\\
    &+\frac{2}{3}R_{XYXY}XYdXdY,
\end{align}
where all the quantities are expressed in terms of the Fermi coordinates obtained in Section~\ref{Fermi coordinates}. \\
The coefficients $R_{0i0j}$ correspond to the gravitoelectric part of the curvature~\cite{Ni:1978zz}, associated with tidal accelerations between neighboring free-falling particles, as expressed by the geodesic deviation equation. They thus describe stretching or compression of the locally inertial congruence. The mixed terms $R_{0ijk}$ correspond to the gravitomagnetic part of the curvature, encoding rotational effects. They govern the precession of gyroscopes, and thus of locally inertial spatial axes. In the standard weak-field, slow-motion limit, they reduce to spatial derivatives of the gravitomagnetic field, providing the fully relativistic generalization of the Lense-Thirring effect (frame dragging). Finally, to the second order in the Fermi coordinates, $g_{ij}$ represents the metric induced on the observer's rest space. The terms $R_{ijkl}$ thus describe the intrinsic curvature of these local spatial hypersurfaces, quantifying the deviations from Euclidean geometry experienced by the observer's laboratory frame. \\
The expression~\eqref{Fermi metric} can be specified to the case of rigidly rotating dust, yielding 
\begin{align}
\label{BG metric Fermi coordinates}
    ds^{2}|_{BG,\,Z=0}=&-[1+R_{TXTX}X^{2}+R_{TYTY}Y^{2}]dT^{2} \notag \\
    &-\frac{4}{3}[R_{TXXY}XY+R_{TYXY}Y^{2}]dTdX \notag\\
    &-\frac{4}{3}[R_{TXXY}X^{2}+R_{TYXY}XY]dTdY \notag \\
    &+\left[1-\frac{1}{3}R_{XYXY}Y^{2}\right]dX^{2}+\left[1-\frac{1}{3}R_{XYXY}X^{2}\right]dY^{2} \notag\\
    &+\frac{2}{3}R_{XYXY}XYdXdY.
\end{align}
This construction completes the characterization of the locally inertial system associated with an observer following a dust geodesic in the galactic plane of stationary axisymmetric dust solutions of the EE without the need to extend them to asymptotic regions that lie outside their domain of validity when applied to model disc galaxies.

%%%%%%%%%%%%%%%%%%%%%%%%%%%%%%%%%%%%%%%%%%%

\subsection{Radially locked reference system}
\label{radially locked reference system}

To interpret spectroscopic and astrometric measurements of photons coming from distant regions of a galaxy, or from other galaxies, one must compare observations made by different locally inertial observers. This requires a consistent prescription for relating their local reference frames. Since each locally inertial tetrad is constructed along a single worldline and is valid only within a limited neighborhood, a rule for fixing their relative orientation is necessary to build a coherent description across the spacetime.

In practice, physical quantities are first computed in the locally inertial system adapted to the observer performing the measurement, its proper laboratory system, and then transformed to a coordinate system sharing a common, operationally defined direction. In the BCRS the axes are tied to distant non-rotating sources (see Section~\ref{BCRS}). This approach, however, suffers from a fundamental limitation: such sources lie outside the spacetime domain described by the general relativistic dust models of galaxies, and the static coordinates themselves may lose meaning depending on the external matter distribution. To overcome this limitation, we seek a purely local prescription allowing different observers to share a common spatial direction. A natural choice is to align one axis with a radial direction defined locally by light propagation. This direction differs from the coordinate vector $\partial_{r}$ used in the BCRS because of frame dragging, and it is defined instead by radial null geodesics with zero angular momentum passing through the galactic center. This construction defines what we shall call the radially locked reference system. \\
Setting $\epsilon=0$ and $L=0$ in the geodesic equations~\eqref{integrated geodesic eq t},~\eqref{integrated geodesic eq phi},~\eqref{integrated geodesic eq r,z} gives
\begin{subequations}
\begin{align}
    \dot{t}&=e^{-2\nu}E, \\
    \dot{\phi}&=e^{-2\nu}\chi E, \\
    \dot{r}&=e^{-\nu-\mu/2}\lvert E\rvert.
\end{align}
\end{subequations}
This definition is meaningful only if these equations admit a unique solution under suitable boundary conditions. In particular, photons emitted from the galactic center must reach the spacetime point of interest along a unique trajectory. Local existence and uniqueness follow from the Picard–Lindelöf theorem~\cite{Hartman} if the metric derivatives satisfy a local Lipschitz condition. Global uniqueness requires a global Lipschitz condition, satisfied when the derivatives of the metric remain bounded throughout spacetime. Physically relevant metrics are $C^{\infty}$ and often analytic, so global Lipschitz continuity holds provided no singularities are present~\cite{Hawking1973,Stephani:2003tm}. In $(\eta,H)$ spacetimes with asymptotic flatness, potential singularities may occur along the rotation axis. Imposing regularity there guarantees global existence and uniqueness of the radial geodesics. Specifically, near $r=0$ we require $g_{\phi\phi}\to r^{2}$, which holds if $\eta\simeq r^{2}$, and finiteness of the angular velocity $\Omega\simeq\Omega_{0}$, which implies $H\simeq H_{0}+ar^{3}$. Under these conditions, the galactic plane is regular and the metric components analytic, so the geodesic equations are globally Lipschitz continuous. The global version of the Picard–Lindelöf theorem then ensures the uniqueness of radial null geodesics for each value of $E$. The tangent vector to such geodesics is 
\begin{equation}
\label{k global coordinates}
    k=Ee^{-\nu}(e^{-\nu}\partial_{t}+e^{-\nu}\chi\partial_{\phi}+e^{-\mu/2}\textrm{sign}(E)\partial_{r}).
\end{equation}
Performing the change of basis as in~\eqref{change of basis t},~\eqref{change of basis phi},~\eqref{change of basis r}, it becomes
\begin{align}
\label{four-momentum radial photon}
    k=\frac{E}{\sqrt{-H}}&\left\{e_{0}+\frac{1}{r}\left[(-\eta\sin{(\omega\tau)}+\sqrt{r^{2}-\eta^{2}}\textrm{sign}(E)\cos{(\omega\tau)})e_{2}\right.\right. \notag \\
    &\left.\left.-(\eta\cos{(\omega\tau)}+\sqrt{r^{2}-\eta^{2}}\textrm{sign}(E)\sin{(\omega\tau)})e_{3}\right]\right\}.
\end{align}
Projecting $k$ onto the rest space of the local observer via $P_{u}=g+u\otimes u$ yields the spatial direction measured in the observer’s frame,
\begin{align}
    k_{obs}=P_{u}k=&E\left[\left(e^{-2\nu}+\frac{1}{H}\right)\partial_{t}+\left(e^{-2\nu}\chi+\frac{\Omega}{H}\right)\partial_{\phi} \right. \notag \\
    &\left.+e^{-\nu-\mu/2}\textrm{sign}(E)\partial_{r}\right].
\end{align}
Performing a change of basis to write the line of sight in the locally inertial reference system gives
\begin{align}
    k_{obs}&=\left[\frac{-\eta}{r\sqrt{-H}}\sin{(\omega\tau)}+e^{-2\nu}\textrm{sign}(E)\cos{(\omega\tau)}\right]e_{2} \notag \\
    &+\left[\frac{-\eta}{r\sqrt{-H}}\cos{(\omega\tau)}+e^{-2\nu}\textrm{sign}(E)\sin{(\omega\tau)}\right]e_{3},
\end{align}
where everything is expressed in Fermi coordinates $(T,X,Y)$, or equivalently $(T,s,\theta)$. The vector $k_{obs}$ needs to be normalized to obtain the unit vector pointing in the observed radial direction $\hat{k}_{obs}$. In particular, we impose $\hat k^{2}_{obs}=1$, while $k^{2}_{obs}=\eta_{ab}k^{a}_{obs}k^{b}_{obs}=-\frac{1}{H}$.
Therefore,
\begin{align}
    \hat{k}_{obs}=&\sqrt{-H}k_{obs}= \\ \notag 
    =&\left(-\frac{\eta}{r}\sin{(\omega\tau)}+\frac{\sqrt{r^{2}-\eta^{2}}}{r}\textrm{sign}(E)\cos{(\omega\tau)}\right)e_{2} \notag \\
    &+\left(-\frac{\eta}{r}\cos{(\omega\tau)}-\frac{\sqrt{r^{2}-\eta^{2}}}{r}\textrm{sign}(E)\sin{(\omega\tau)}\right)e_{3}.
\end{align}
Finally, we perform a Lorentz transformation, namely a rotation in the $e_{2}-e_{3}$ plane, to bring the spacelike vector $e_{2}$ in the radial null direction. Let us call the transformed tetrad $\{e_{0}',e_{1}',e_{2}',e_{3}'\}$. Then,
\begin{equation}
    e_{a}'=\Lambda_{a}^{b}e_{b},
\end{equation}
where
\begin{equation}
    \Lambda=
    \begin{pmatrix}
        1 & 0 & 0 & 0 \\
        0 & 1 & 0 & 0 \\
        0 & 0 & \cos{\theta(\tau)} & \sin{\theta(\tau)} \\
        0 & 0 & -\sin{\theta(\tau)} & \cos{\theta(\tau)}
    \end{pmatrix}.
\end{equation}
Imposing $e_{2}'=\hat{k}_{obs}$ we obtain 
\begin{equation}
    \cos{\theta(\tau)}=\frac{1}{r}[(-\eta\sin{(\omega\tau)}+\sqrt{r^{2}-\eta^{2}}\textrm{sign}(E)\cos{(\omega\tau)})],
\end{equation}
\begin{equation}
    \sin{\theta(\tau)}=-\frac{1}{r}[(\eta\cos{(\omega\tau)}+\sqrt{r^{2}-\eta^{2}}\textrm{sign}(E)\sin{(\omega\tau)})].
\end{equation}
This yields a tetrad ${e_{a}'}$ adapted to a geodesic observer with a spatial axis pointing along the radial null direction, defining the radially locked frame:
\begin{subequations}
\begin{align}
    e_{0}'(\tau)=&\frac{1}{\sqrt{-H}}(\partial_{t}+\Omega\partial_{\phi}), \\
    e_{1}'(\tau)=&e^{-\frac{\mu}{2}}\partial_{z}, \\
    e_{2}'(\tau)=&\frac{1}{r}[(-\eta\sin{(\omega\tau)}+\sqrt{r^{2}-\eta^{2}}\textrm{sign}(E)\cos{(\omega\tau)})]e_{2} \notag \\
    &-\frac{1}{r}[(\eta\cos{(\omega\tau)}+\sqrt{r^{2}-\eta^{2}}\textrm{sign}(E)\sin{(\omega\tau)})]e_{3}, \\
    e_{3}'(\tau)=&\frac{1}{r}[(\eta\cos{(\omega\tau)}+\sqrt{r^{2}-\eta^{2}}\textrm{sign}(E)\sin{(\omega\tau)})]e_{2} \notag \\
    &+\frac{1}{r}[(-\eta\sin{(\omega\tau)}+\sqrt{r^{2}-\eta^{2}}\textrm{sign}(E)\cos{(\omega\tau)})]e_{3}.
\end{align}
\end{subequations}
Although this reference system is obtained by a time-dependent rotation of the locally inertial tetrad, and is therefore not inertial, it provides a physically meaningful way to align the orientation of neighboring laboratory frames through the propagation of light. Its orientation is fixed by local, operationally measurable quantities and provides a kinematically non-rotating reference system without the need to rely on distant or asymptotic reference objects. It thus represents the general-relativistic analogue of a celestial coordinate system constructed from the intrinsic geometry of light propagation, independent of any external background.  \\
Finally, one may envisage generalizing this construction to larger configurations containing multiple galaxies. Once a procedure to glue together different galactic spacetimes is prescribed, the corresponding radially locked frames could be consistently aligned across the intervening space. This would allow for a meaningful comparison of measurements performed by locally inertial observers in different galaxies, without invoking asymptotic inertial directions or distant fixed sources.

%%%%%%%%%%%%%%%%%%%%%%%%%%%%%%%%%%%%%%%%%%%%%

\section{Frequency shift}
\label{frequency shift}

A central observable in galactic dynamics is the frequency shift of photons exchanged between different regions of the system. This quantity is typically used to compute the radial velocity of stars in galaxies and underlies the definition of spectroscopic velocity~\cite{LindegrenDravins2003}. In GR, frequency shifts encode not only the kinematical Doppler effect associated with the relative motion between the emitter and the receiver, but also purely gravitational effects arising from spacetime curvature and frame dragging~\cite{MisnerThorneWheeler1973,Wald:1984}. Its computation therefore provides a direct means of distinguishing general relativistic models of disc galaxies from their Newtonian counterparts~\cite{Astesiano:2022tbh}. \\
Consider the ideal situation in which a photon is emitted by a star with frequency $\nu_{e}$, as measured by a locally inertial observer attached to it, and propagates along a radial light ray until it reaches another star, where a second observer measures the frequency $\nu_{o}$ in its locally inertial reference system. The affine parameter along the lightlike geodesic with tangent vector $k$ can always be chosen such that $\nu=-g(u,k)$ yields the photon frequency as measured by an observer with four-velocity $u$. Then, the frequency measured by the locally inertial observer at the emission event can be computed as 
\begin{equation}
    \nu_{e}=-g_{\mu\nu}u_{e}^{\mu}k^{\nu},
\end{equation}
where $g_{\mu\nu}$ is the metric in a neighborhood of the emitting star geodesic in Fermi normal coordinates~\eqref{BG metric Fermi coordinates}, $u_{e}=e_{0}$ is the observer's four-velocity in its own reference system, and $k$ is the tangent vector to the light ray. Restricting to the galactic plane, the latter is obtained by setting $\epsilon=0$ and $\dot z=0$ in equations~\eqref{integrated geodesic eq t},~\eqref{integrated geodesic eq phi},~\eqref{integrated geodesic eq r,z},
\begin{align} 
        k=e^{-2\nu}(E-\chi L)\partial_{t}+[\chi e^{-2\nu}(E-\chi L)+Le^{-2\psi}]\partial_{\phi}\pm e^{-\frac{\mu}{2}}[e^{-2\nu}(e-\chi L)^{2}+e^{-2\psi}L^{2}]^{\frac{1}{2}}\partial_{r},
\end{align}
where $E$ and $L$ are the photon conserved energy and angular momentum, respectively.
Performing the change of basis to the locally inertial system as in~\eqref{change of basis t},~\eqref{change of basis phi},~\eqref{change of basis r}, it becomes
\begin{align}
\label{general lightlike geodesic tangent vector}
    k=&\frac{1}{\sqrt{-H}}(E-\Omega L)e_{0} \notag \\
    +&\left\{\frac{1}{\sqrt{-H}r}[(E-\Omega L)\eta-HL]\sin{(\omega\tau)}\pm\left[-\frac{HL^{2}}{r^{2}-\eta^{2}}-\frac{r^{2}-\eta^{2}}{Hr^{2}}\left(E-\Omega L-\frac{H\eta}{\eta^{2}-r^{2}}L\right)^{2}\right]^{\frac{1}{2}}\cos{(\omega\tau)}\right\}e_{2}\notag \\
    +&\left\{-\frac{1}{\sqrt{{-H}}r}[(E-\Omega L)\eta-HL]\cos{(\omega\tau)}\mp\left[-\frac{HL^{2}}{r^{2}-\eta^{2}}-\frac{r^{2}-\eta^{2}}{Hr^{2}}\left(E-\Omega L-\frac{H\eta}{\eta^{2}-r^{2}}L\right)^{2}\right]^{\frac{1}{2}}\sin{(\omega\tau)}\right\}e_{3}.
\end{align}
Therefore we have
\begin{align}
\label{emission frequency LIO}
    \nu_{e}&=(-g_{00}u_{e}^{0}k^{0}-g_{02}u_{e}^{0}k^{2}-g_{03}u_{e}^{0}k^{3})_{e} \notag \\
    &=[1+R_{0202}(X_{e}^{2})^{2}+2R_{0203}X_{e}^{2}X_{e}^{3}+R_{0303}(X_{e}^{3})^{2}]\frac{1}{\sqrt{-H_{e}}}(E-\Omega_{e} L) \notag \\
    & +\frac{2}{3}[R_{0223}X_{e}^{2}X_{e}^{3}+R_{0323}(X_{e}^{3})^{2}] \notag \\
    &\times\left\{\frac{1}{\sqrt{-H_{e}}r_{e}}[(E-\Omega_{e} L)\eta_{e}-H_{e}L]\sin{(\omega_{e}\tau_{e})}\pm\left[-\frac{H_{e}L^{2}}{r_{e}^{2}-\eta_{e}^{2}}-\frac{r_{e}^{2}-\eta_{e}^{2}}{H_{e}r_{e}^{2}}\left(E-\Omega_{e} L-\frac{H_{e}\eta_{e}}{\eta_{e}^{2}-r_{e}^{2}}L\right)^{2}\right]^{\frac{1}{2}}\cos{(\omega_{e}\tau_{e})}\right\} \notag  \\
    &+\frac{2}{3}[R_{0223}(X_{e}^{2})^{2}+R_{0323}X_{e}^{2}X_{e}^{3}] \notag \\
    &\times\left\{-\frac{1}{\sqrt{{-H_{e}}}r_{e}}[(E-\Omega_{e} L)\eta_{e}-H_{e}L]\cos{(\omega_{e}\tau_{e})}\mp\left[-\frac{H_{e}L^{2}}{r_{e}^{2}-\eta_{e}^{2}}-\frac{r_{e}^{2}-\eta_{e}^{2}}{H_{e}r_{e}^{2}}\left(E-\Omega_{e} L-\frac{H_{e}\eta_{e}}{\eta_{e}^{2}-r_{e}^{2}}L\right)^{2}\right]^{\frac{1}{2}}\sin{(\omega_{e}\tau_{e})}\right\},
\end{align}
where the subscripts indicate that all the functions are evaluated at the emission event $X^{\mu}_{e}$. \\
An analogous expression holds at the observation event, with all quantities evaluated there. If both the emitter and the observer are dust particles following geodesics, then the emission and observation events occur on their respective trajectories, where the locally inertial metric reduces to the Minkowski one. In this case, $X^{\mu}_{e}=0$ and $X^{\mu}_{o}=0$, so that
\begin{equation}
    \nu_{e/o}=\frac{1}{\sqrt{-H_{e/o}}}(E-\Omega_{e/o}L).
\end{equation}
Therefore, as the photon moves from the emission to the observation point along its worldline, it experiences a frequency shift that can be expressed through the redshift factor
\begin{equation}
\label{redshift factor}
    z=\frac{\nu_{o}}{\nu_{e}}-1=\sqrt{\frac{H_{e}}{H_{o}}}\frac{E-\Omega_{o}L}{E-\Omega_{e}L}-1.
\end{equation}
In a rigidly rotating dust model, such as the BG model, $\Omega=\text{const.}$ and $H=-1$. As a consequence, also the combination $E-\Omega L$ is constant along the photon geodesic, implying that photons exchanged between different rigidly rotating dust elements exhibit no redshift. \\
From a kinematical perspective, the absence of redshift follows from the way local inertial frames are related along the null geodesic. The frequency measured by each observer is determined by the time component of the photon wavevector in their local rest frame. A non-vanishing redshift arises only if the emitter's local frame, when parallel transported along the light ray, is Lorentz-boosted with respect to the receiver's local frame. In the case of rigid rotation, the parallel transport of the emitter's tetrad along the null geodesic preserves the alignment of the time axes. As a consequence, the photon wavevector has identical time components in the two local frames, yielding a vanishing redshift. We stress that such a vanishing redshift does not imply vanishing relative motion between the emitter and the observer, since the interpretation of a frequency shift in terms of a radial velocity requires additional assumptions on the transverse motion.

%%%%%%%%%%%%%%%%%%%%%%%%%%%%%%%%%%%%%%%

\section{Reconstructing relative velocities from observations in a locally inertial reference system}
\label{calculation of relative velocity}

We now outline an operational procedure to reconstruct the relative velocities of a dust element in a $(\eta,H)$ spacetime with respect to a locally inertial observer, directly from spectroscopic and astrometric observables. Such an observer moves around the galactic center within a spacetime that describes the averaged motion of a portion of the disc, and does not coincide with the worldline of any specific star or satellite. These, in general, do not follow circular trajectories in a stationary axisymmetric spacetime but have their own peculiar motions with respect to such a background. Consequently, in order to extract rotation curves, one must also account for the motion of the observer relative to a non-rotating reference system. To this end, one would have to identify a representative dust element whose motion anchors that of a real observer to the effective model. \\
Rotation curves, which express radial velocities of stars as a function of the distance from the galactic center, are originally Newtonian constructs. Their definition relies on the existence of global inertial reference systems and on Galilean addition rules to translate Doppler shifts observed in the Solar System into velocities referred to the galactic center~\cite{SofueRubin2001}. In this framework, the radial velocity is the object's motion along the line of sight, as deduced from spectral line shifts. 
This concept was reformulated within the BCRS framework in the IAU 2000 resolutions~\cite{LindegrenDravins2003}. There, the barycentric radial velocity measure $cz_{B}$ was introduced as the outcome of a spectroscopic measurement, with $z_{B}$ denoting the redshift reduced to the Solar System barycenter. This definition avoids identifying $cz_{B}$ with a physical radial velocity. Translating the spectroscopic barycentric radial velocity measure into the physical velocity of the object is model-dependent since $cz_{B}$ is just a quantification of the spectroscopic shift, not a physical velocity. The interpretation of $cz_{B}$ as a spectroscopic radial velocity relies on the full description of the astronomical event, consisting of the motion of the emitting star, the propagation of light through the observer, the motion of the observer, and the measurement of the signal. Moreover, since the spectroscopic velocity measure is defined in the BCRS, it inherits the limitations associated with that reference system. In the following, we propose a method to obtain a physical radial spectroscopic velocity as measured by a locally inertial observer in a stationary axisymmetric dust solution of the EE. \\
As discussed in Section~\ref{relative velocities}, reconstructing $v_{spec}^{rad}$ from a measured redshift is not trivial, since in a generic configuration the redshift depends on both the radial and tangential components of the relative motion.
For distant galaxies and stars, however, the contribution of the transverse motion enters only at second order in $v/c$. Under these conditions, and assuming an approximately Minkowskian background, the total frequency shift is dominated by the radial term, so that $v_{spec}\sim v_{spec}^{rad}$, and the latter can be obtained directly from the redshift using Eq.~\eqref{velocity from redshift}. In the general relativistic dust models considered here, however, both the observer and the source move on circular orbits within the galactic plane, so the radial and tangential components of the spectroscopic velocity are of comparable magnitude. In this regime, the redshift alone does not allow one to isolate $v_{spec}^{rad}$. \\
Astrometry provides the complementary information by measuring the rate of change of the apparent direction of the incoming light ray with respect to the observer's locally inertial tetrad. In practice, this quantity can be obtained from high-precision satellite missions such as Gaia, which provide the measure of the direction cosines of the photon’s spatial momentum with respect to the axes of a local reference system adapted to the satellite~\cite{Crosta2019}:
\begin{equation}
    \cos{\psi_{e_{a},\hat{k}_{obs}}}=\frac{P(u)_{\alpha\beta}k^{\alpha}e^{\beta}_{a}}{[P(u)_{\alpha\beta}k^{\alpha}k^{\beta}]^{\frac{1}{2}}}, \quad a=1,2,3,
\end{equation}
where $P(u)=g+u\otimes u$ projects onto the observer's rest space, and $k$ is the tangent vector to the photon's geodesic. Specializing to the galactic plane and using the locally inertial reference system defined previously, with $k$ given by Eq.~\eqref{general lightlike geodesic tangent vector}, one obtains
\begin{equation}
    \cos{\psi_{e_{a},\hat{k}_{obs}}}=P(u)_{\alpha\beta}\hat{k}^{\alpha}e^{\beta}_{a}=\eta_{\alpha\beta}\hat{k}^{\alpha}_{obs}e^{\beta}_{a}.
\end{equation}
Therefore,
\begin{equation}
    \cos{\psi_{e_{2},\hat{k}_{obs}}}=\frac{1}{\sqrt{-H}r}[(E-\Omega L)\eta-HL]\sin{(\omega\tau)}\pm\left[-\frac{HL^{2}}{r^{2}-\eta^{2}}-\frac{r^{2}-\eta^{2}}{Hr^{2}}\left(E-\Omega L-\frac{H\eta}{\eta^{2}-r^{2}}L\right)^{2}\right]^{\frac{1}{2}}\cos{(\omega\tau)},
\end{equation}
\begin{equation}
    \cos{\psi_{e_{3},\hat{k}_{obs}}}=-\frac{1}{\sqrt{{-H}}r}[(E-\Omega L)\eta-HL]\cos{(\omega\tau)}\mp\left[-\frac{HL^{2}}{r^{2}-\eta^{2}}-\frac{r^{2}-\eta^{2}}{Hr^{2}}\left(E-\Omega L-\frac{H\eta}{\eta^{2}-r^{2}}L\right)^{2}\right]^{\frac{1}{2}}\sin{(\omega\tau)}.
\end{equation}
Differentiating these angles with respect to the observer's proper time gives the apparent proper motion of the source and therefore the astrometric relative velocity. \\
To reconstruct the full relative velocity of a dust particle with respect to a locally inertial observer, the spectroscopic and astrometric velocities must be combined.  
In general curved spacetimes, there is no exact algebraic relation between $v_{spec}$ and $v_{ast}$. However, within a Fermi neighborhood around the observer's geodesic, the metric is Minkowskian up to quadratic tidal corrections, and to leading order, one can employ the flat space relation
\begin{equation}
\label{relation between vspec and vast}
    v_{ast}=\frac{1}{1+g\left(v_{spec},\frac{s_{obs}}{||s_{obs}||}\right)}v_{spec},
\end{equation}
where $s_{obs}$ is the relative position of the source relative to the observer, and the observer is geodetic. Combining this expression with the redshift formula~\eqref{velocity from redshift} then yields the radial spectroscopic velocity of the dust with respect to the locally inertial observer. \\
The validity of this reconstruction is limited by the range of the Fermi expansion, i.e.
\[
\lvert R_{\alpha\beta\mu\nu}\rvert \lVert s_{\mathrm{obs}}\rVert^{2} \ll 1,
\]
so that the error induced by using the flat-space relation remains of order $\mathcal{O}(\lvert R_{\alpha\beta\mu\nu}\rvert \lVert s_{\mathrm{obs}}\rVert^{2})$. A more accurate treatment would require including curvature corrections to Eq.~\eqref{relation between vspec and vast}, which can, in principle, be computed from the expansion of the photon geodesic in Fermi coordinates. An equivalent construction can be performed after transforming to the radially locked reference system introduced in Section~\ref{radially locked reference system}, where the common orientation of spatial axes simplifies the comparison of velocities measured by different local observers. \\
With these ingredients, once the radial component of the spectroscopic velocity has been reconstructed from combined redshift and astrometric data, one may, in principle, construct a general relativistic analogue of a rotation curve. This is achieved by consistently combining the motion of the actual observer performing the measurement relative to the locally inertial observer attached to a representative dust element, with the motion of that locally inertial observer relative to the galactic center. This, as well as a full treatment of proper-motion reconstruction and curvature corrections along null geodesics, will be the subject of subsequent work; our aim here is to set up the locally inertial framework and clarify how, in principle, observational procedures map onto the GR notions of relative velocity.

%%%%%%%%%%%%%%%%%%%%%%%%%%%%%%%%%%%%%%%%%%%%%%

\section{Conclusions}
\label{conclusions}

In this work, we have constructed a locally inertial reference system adapted to geodesic observers in stationary, axisymmetric self-gravitating dust solutions of the EE. Within this framework, we have outlined a consistent procedure to reconstruct relative velocities of the dust with respect to such observers using spectroscopic and astrometric observables. The guiding principle was to account for the intrinsically local character of general relativistic galactic models, without relying on post-Newtonian approximations that treat the Solar System as isolated or on the existence of distant non-rotating reference objects. To this end, we adopted locally inertial observers, which provide the closest relativistic analogue of Newtonian inertial frames, corresponding to the local laboratory frames in which measurements are actually performed. The associated reference system is constructed by identifying the timelike axis with the tangent vector to the observer's worldline and defining the spatial axes through Fermi–Walker transport along dust geodesics in the galactic plane. We then obtained the corresponding Fermi coordinates and the explicit form of the metric in a neighborhood of the observer's trajectory. Unlike the BCRS and ZAMO constructions, this procedure yields a dynamically non-rotating reference system, whose spatial axes are operationally set up by a triad of mutually orthogonal gyroscopes transported along the observer's worldline. \\
To enable different locally inertial observers to consistently compare the orientation of their spatial axes, we introduced the radially locked reference system, in which one spatial axis is aligned with the radial direction defined by photons passing through the galactic center with zero angular momentum. Unlike the BCRS, which relies on the existence of distant objects with negligible proper motion, this construction yields a kinematically non-rotating reference system that can be established entirely locally by orienting one spatial axis along the observed direction connecting the dust observer to the galactic center. \\
Having established the reference system, we then examined how such observers describe a $(\eta,H)$ galaxy starting from the frequency shift of photons exchanged between two points. We considered the idealized configuration in which a photon is emitted with a given frequency measured in the locally inertial frame comoving with the emitting dust particle, propagates along a null geodesic, and is subsequently received by another dust particle, where its frequency is measured in the corresponding locally inertial frame. As expected, the associated redshift vanishes for rigidly rotating dust configurations, while it is generically nonzero in the presence of differential rotation. In GR, frequency shifts encode not only the kinematical Doppler contribution arising from the relative motion of emitter and receiver, but also genuinely gravitational effects due to spacetime curvature and frame dragging. Their explicit computation, therefore, provides an opportunity to test the need for a general relativistic description of the spacetime. In observational practice, redshift measurements are used within the BCRS to define the radial velocity measure that underlies the construction of galactic rotation curves. Interpreting the redshift as a physical radial velocity, however, requires modeling the full observational process and, in general, cannot be achieved from the frequency shift alone without additional assumptions or approximations. This motivated us to address how relative velocities in GR can be employed to define a spectroscopic radial velocity with respect to locally inertial observers. To this end, we proposed a procedure that combines spectroscopic and astrometric measurements performed in the observer’s locally inertial frame, yielding a physically meaningful radial velocity within stationary, axisymmetric self-gravitating dust solutions. \\
Much work remains to assess this proposal in practice. A central aspect concerns the use of a stationary axisymmetric dust solution of the EE as an effective model of a portion of the galactic disc. Such solutions are understood as describing an average motion in regions where their underlying assumptions are expected to hold, while necessarily neglecting the peculiar motions of individual disc constituents. As a consequence, the locally inertial reference system constructed here, together with the associated radially locked frame, are adapted to idealized dust geodesics that do not, in general, coincide with the trajectory of an actual observational platform. A key step for the practical application of this framework to the reconstruction of galactic rotation curves is therefore the identification of a suitable reference dust worldline that can serve as a representative anchor between the effective stationary, axisymmetric spacetime and the motion of a real observer, such as Gaia. While the present construction has the advantage of not relying on asymptotically defined ``fixed stars'' and instead refers only to the galactic center and to a local dust congruence, the explicit operational identification of such a reference dust element remains open. At present, the motion of an actual observer can still only be characterized relative to distant sources, and bridging this description with the effective dust geometry adopted here is an essential ingredient for direct comparison with observational data. \\
From a more technical perspective, an explicit implementation of the proposed strategy also requires the detailed computation of the spectroscopic radial velocity relative to locally inertial observers. This entails a careful reconstruction of proper motions and their consistent combination with redshift measurements within the observer's local neighborhood. Completing these steps would make it possible to extract velocity profiles calculated in a fully consistent general relativistic framework that could be directly confronted with observed galactic rotation curves.

%%%%%%%%%%%%%%%%%%%%%%%%%%%%%%%%%%%%%%%%%%%%%

\appendix

\section{Appendix A}
\label{A: Gyroscope equation solution}
In this appendix we derive explicitly the solution to the gyroscope equation~\eqref{gyroscope equations matrix form}:
\begin{align}
 \dot S=MS,
\end{align}
where $M$ is the constant matrix
\begin{align}
 M=
\begin{pmatrix}
 0 & \alpha & \beta & 0 \\ 
 \alpha \frac \Omega{\eta} r^2 e^{-\mu} & 0 & 0 & -\alpha \frac {r^2}\eta e^{-\mu} \\
 \beta \frac \Omega{\eta} r^2 e^{-\mu} & 0 & 0 & -\beta \frac {r^2}\eta e^{-\mu} \\ 
 0 & \alpha (\Omega-\frac H\eta) & \beta (\Omega-\frac H\eta) & 0
\end{pmatrix},\label{M matrix}
\end{align}
with
\begin{align}
 \alpha=& \frac {\nu,_z}{2\sqrt {-H}} \left( 1-\frac {\eta^2}{r^2} \right), \\
 \beta=& \frac {1}{2\sqrt {-H}} \left( \nu,_r( 1-\frac {\eta^2}{r^2} ) +\frac {\eta^2}{r^3} \right).
\end{align}
Since all matrix elements are independent from $\tau$ (being evaluated at constant $r$ and $z$), the solution of the equation is simply
\begin{align}
 S(\tau)=e^{M \tau} S_0,
\end{align}
where 
\begin{align}
 S_0=\begin{pmatrix}
 S^t_0 \\ S^z_0 \\ S^r_0 \\ S^\phi_0
\end{pmatrix},
\end{align}
is the initial configuration (at $\tau=0$). \\
In order to compute $e^{M \tau}$, it is convenient to write
\begin{align}
 M=R D R^{-1},
\end{align}
where $D$ is the diagonal form of $M$, so that
\begin{align}
 e^{M \tau}=R e^{D\tau}R^{-1}.
\end{align}
The eigenvalues of $M$ can be determined as follows. Notice that since the first and last rows of $M$ are linearly dependent, the matrix has one zero eigenvalue. Similarly, the second and third rows are linearly dependent, yielding a second zero eigenvalue. As the trace of $M$ vanishes, the remaining two eigenvalues must be one the opposite of the other. In particular, they satisfy
\begin{align}
 2\lambda^2=\textit{Tr} M^2=2(\alpha^2+\beta^2)(x-yz),
\end{align}
where we have introduced
\begin{align}
 x=\frac \Omega{\eta} r^2 e^{-\mu}, \qquad y=\frac {r^2}\eta e^{-\mu} , \qquad  z= \Omega-\frac H\eta.
\end{align}
Observe that 
\begin{align}
 x-yz=\frac {r^2}{\eta^2} e^{-\mu} H
\end{align}
is negative, so that $\lambda$ is imaginary. The four eigenvalues are then
\begin{align}
 0^2, \qquad \pm \frac i2 \frac r\eta e^{-\frac \mu2} \sqrt {\nu,_z^2 \left( 1-\frac {\eta^2}{r^2} \right)^2+\left( \nu,_r\left( 1-\frac {\eta^2}{r^2} \right) +\frac {\eta^2}{r^3} \right)^2}.
\end{align}
The eigenvectors corresponding to the $0$ eigenvalues are
\begin{align}
 v_1=
\begin{pmatrix}
 y \\ 0 \\ 0 \\ x
\end{pmatrix}=\frac {r^2}\eta e^{-\mu}
\begin{pmatrix}
 1 \\ 0 \\ 0 \\ \Omega
\end{pmatrix},\qquad
 v_2=
\begin{pmatrix}
 0 \\ -\beta \\ \alpha \\ 0
\end{pmatrix}=-\frac {1}{2\sqrt {-H}}
\begin{pmatrix}
 0 \\  \nu,_r\left( 1-\frac {\eta^2}{r^2} \right) +\frac {\eta^2}{r^3}  \\ -\nu,_z \left( 1-\frac {\eta^2}{r^2} \right) \\ 0
\end{pmatrix},
\end{align}
Moreover, if $v_3$ is the eigenvector corresponding to $\lambda$, then $v_4=\bar v_3$ is its complex conjugate. A direct calculation gives
\begin{align}
 v_3=
\begin{pmatrix}
 \sqrt {\alpha^2+\beta^2} \\ i\alpha \sqrt {yz-x} \\ i\beta \sqrt {yz-x} \\ z\sqrt {\alpha^2+\beta^2}
\end{pmatrix}.
\end{align}
Therefore, the diagonalizing matrix is
\begin{align}
 R=(v_1v_2v_3v_4)=
\begin{pmatrix}
 y & 0 & \sqrt {\alpha^2+\beta^2} & \sqrt {\alpha^2+\beta^2} \\
 0 & -\beta & i\alpha \sqrt {yz-x} & -i\alpha \sqrt {yz-x} \\
 0 & \alpha & i\beta \sqrt {yz-x} & -i\beta \sqrt {yz-x} \\
 x & 0 & z\sqrt {\alpha^2+\beta^2} & z\sqrt {\alpha^2+\beta^2}
\end{pmatrix}.
\end{align}
The inverse matrix is
\begin{align}
 R^{-1} =
\begin{pmatrix}
 \frac z{yz-x} & 0 & 0 & -\frac 1{yz-x}\\
 0 & -\frac \beta{\alpha^2+\beta^2} & \frac \alpha{\alpha^2+\beta^2} & 0\\
 -\frac x2 \frac 1{\sqrt {\alpha^2+\beta^2}(zy-x)} & -\frac {i\alpha}{2(\alpha^2+\beta^2)\sqrt{yz-x}} & -\frac {i\beta}{2(\alpha^2+\beta^2)\sqrt{yz-x}}  & \frac y2 \frac 1{\sqrt {\alpha^2+\beta^2}(zy-x)}\\
 -\frac x2 \frac 1{\sqrt {\alpha^2+\beta^2}(zy-x)} & \frac {i\alpha}{2(\alpha^2+\beta^2)\sqrt{yz-x}} & \frac {i\beta}{2(\alpha^2+\beta^2)\sqrt{yz-x}}  & \frac y2 \frac 1{\sqrt {\alpha^2+\beta^2}(zy-x)}
\end{pmatrix}.
\end{align}
If we conveniently introduce $\omega$ such that $\lambda=i\omega$,
\begin{align}
 \omega=\frac 12 \frac r\eta e^{-\frac \mu2} \sqrt {\nu,_z^2 \left( 1-\frac {\eta^2}{r^2} \right)^2+\left( \nu,_r\left( 1-\frac {\eta^2}{r^2} \right) +\frac {\eta^2}{r^3} \right)^2},
\end{align}
we finally get
\begin{align}
 e^{M\tau}=R 
\begin{pmatrix}
 1 & 0 & 0 & 0\\
 0 & 1 & 0 & 0\\
 0 & 0 & e^{i\omega\tau} & 0\\
 0 & 0 & 0 & e^{-i\omega\tau}
\end{pmatrix} R^{-1}.
\end{align}
Substituting the explicit expressions for $R$ and its inverse, we obtain the result~\eqref{solution gyroscope equation}.

%%%%%%%%%%%%%%%%%%%%%%%%%%%%%%%%%%%%%%

\section{Appendix B}
\label{A: curvature quantities}

We report the metric quantities used in this work.

\subsection{Christoffel symbols}

The Christoffel symbols can be obtained directly from the definition
\begin{equation}
    \Gamma^{\alpha}_{\mu\nu}=\frac{1}{2}g^{\alpha\beta}(g_{\mu\beta,\nu}+g_{\nu\beta,\mu}-g_{\mu\nu,\beta}):
\end{equation}
Using the metric~\eqref{ZAMO metric} we get for the nonzero coefficients
\begin{align}
 \Gamma^t_{tr}=&\Gamma^t_{rt}=\nu,_r-\frac 12 e^{2(\psi-\nu)}\chi \chi,_r; \qquad \Gamma^t_{tz}=\Gamma^t_{zt}=\nu,_z-\frac 12 e^{2(\psi-\nu)}\chi \chi,_z;\\
 \Gamma^t_{\phi r}=&\Gamma^t_{r\phi}=\frac 12 e^{2(\psi-\nu)} \chi,_r; \qquad \Gamma^t_{\phi z}=\Gamma^t_{z\phi}=\frac 12 e^{2(\psi-\nu)} \chi,_z;\\
 \Gamma^\phi_{\phi r}=&\Gamma^\phi_{r\phi}=\psi,_r+\frac 12 e^{2(\psi-\nu)}\chi \chi,_r; \qquad \Gamma^\phi_{\phi z}=\Gamma^\phi_{z\phi}=\psi,_z+\frac 12 e^{2(\psi-\nu)}\chi \chi,_z;\\
 \Gamma^\phi_{t r}=&\Gamma^\phi_{rt}=-\chi \psi,_r -\frac 12 \chi,_r +\chi\nu,_r -\frac 12 e^{2(\psi-\nu)} \chi^2 \chi,_r; \\  
 \Gamma^\phi_{t z}=&\Gamma^\phi_{zt}=-\chi \psi,_z -\frac 12 \chi,_z +\chi\nu,_z -\frac 12 e^{2(\psi-\nu)} \chi^2 \chi,_z;\\
 \Gamma^r_{zr}=&\Gamma^r_{rz}=\frac 12 \mu,_z; \qquad \Gamma^r_{rr}=\frac 12 \mu,_r; \qquad \Gamma^r_{zz}=-\frac 12 \mu,_r;\qquad  \Gamma^r_{\phi\phi}=-\psi,_r e^{2\psi-\mu};\\
 \Gamma^r_{tt}=&\nu,_r e^{2\nu-\mu} -\chi^2 \psi,_r e^{2\psi-\mu} -\chi \chi,_r e^{2\psi-\mu} ; \qquad \Gamma^r_{\phi t}=\Gamma^r_{t\phi}=(\chi \psi,_r+\frac 12 \chi,_r) e^{2\psi-\mu};\\
 \Gamma^z_{zr}=&\Gamma^z_{rz}=\frac 12 \mu,_r; \qquad \Gamma^z_{zz}=\frac 12 \mu,_z; \qquad \Gamma^z_{rr}=-\frac 12 \mu,_z;\qquad  \Gamma^z_{\phi\phi}=-\psi,_z e^{2\psi-\mu};\\
 \Gamma^z_{tt}=&\nu,_z e^{2\nu-\mu} -\chi^2 \psi,_z e^{2\psi-\mu} -\chi \chi,_z e^{2\psi-\mu} ; \qquad \Gamma^z_{\phi t}=\Gamma^z_{t\phi}=(\chi \psi,_z+\frac 12 \chi,_z) e^{2\psi-\mu}.
\end{align}

\subsection{Ricci's coefficients}

We choose the vierbein forms
\begin{align}
 e^0=e^\nu dt, \qquad e^1=e^{\frac \mu2} dz, \qquad e^2=e^{\frac \mu2} dr, \qquad e^3=e^\psi(d\phi-\chi dt).
\end{align}
The Ricci's coefficients 1-forms $\omega^a_{\ b}$ are defined by the structure equation
\begin{align}
 de^a=-\omega^a_{\ b}\wedge e^b,
\end{align}
together the metric condition $\omega_{ab}=-\omega_{ba}$ (and the indices lowered with the flat Minkowski's metric $\eta_{ab}$ with signature $(-,+,+,+)$). Explicitly:
\begin{align}
 \omega^0_{\ 1} \wedge e^1+\omega^0_{\ 2} \wedge e^2+\omega^0_{\ 3} \wedge e^3=&e^{-\frac \mu2}  \nu,_z e^0\wedge e^1+e^{-\frac \mu2}  \nu,_r e^0\wedge e^2,\\
 \omega^1_{\ 0} \wedge e^0+\omega^1_{\ 2} \wedge e^2+\omega^1_{\ 3} \wedge e^3=&\frac 12 \mu,_r e^{-\frac \mu2} e^1\wedge e^2,\\
 \omega^2_{\ 0} \wedge e^0+\omega^2_{\ 1} \wedge e^1+\omega^2_{\ 3} \wedge e^3=&\frac 12 \mu,_z e^{-\frac \mu2}  e^2\wedge e^1,\\
 \omega^3_{\ 0} \wedge e^0+\omega^3_{\ 1} \wedge e^1+\omega^3_{\ 2} \wedge e^2=&e^{-\frac \mu2}\psi,_z e^3\wedge e^1+e^{-\frac \mu2}\psi,_r e^3\wedge e^2\cr
 &+e^{\psi-\frac \mu2-\nu}\chi,_z e^1\wedge e^0 +e^{\psi-\frac \mu2-\nu}\chi,_r e^2\wedge e^0.
\end{align}
This has solution
\begin{align}
 \omega^0_{\ 1}=&\omega^1_{\ 0}=e^{-\frac \mu2} \nu,_z e^0+\frac 12 e^{\psi-\frac \mu2 -\nu} \chi,_z e^3, \\
 \omega^0_{\ 2}=&\omega^2_{\ 0}=e^{-\frac \mu2} \nu,_r e^0+\frac 12 e^{\psi-\frac \mu2 -\nu} \chi,_r e^3, \\
 \omega^0_{\ 3}=&\omega^3_{\ 0}=\frac 12 e^{\psi-\frac \mu2 -\nu} (\chi,_z e^1+\chi,_r e^2), \\
 \omega^1_{\ 2}=&-\omega^2_{\ 1}=\frac 12 e^{-\frac \mu2} (\mu,_r e^1-\mu,_z e^2), \\
 \omega^1_{\ 3}=&-\omega^3_{\ 1}=\frac 12 e^{\psi-\frac \mu2 -\nu} \chi,_z e^0-e^{-\frac \mu2} \psi,_z  e^3, \\
 \omega^2_{\ 3}=&-\omega^3_{\ 2}=\frac 12 e^{\psi-\frac \mu2 -\nu} \chi,_r e^0-e^{-\frac \mu2} \psi,_r  e^3.
\end{align}

\subsection{Riemann tensor}

Let us start from the computation of the curvature 2-form
\begin{align}
 \Omega^a_{\ b}=d\omega^a_{\ b} +\omega^a_{\ c} \wedge \omega^c_{\ b}. \label{curvform}
\end{align}
It satisfies $\Omega_{ab}=-\Omega_{ba}$. Moreover, if we set
\begin{align}
 \Omega^a_{\ b}=\frac 12  \Omega^a_{\ bcd} e^c \wedge e^d, 
\end{align}
one also has 
\begin{align}
 \Omega_{abcd}=&\Omega_{cdab}, \\
 \Omega_{abcd}=&-\Omega_{abdc}, \\
 \Omega_{abcd}+\Omega_{acdb}+\Omega_{adbc}=&0.
\end{align}
These coefficients are easily calculated directly from~\eqref{curvform}. We get
\begin{align}
 \Omega_{0101}=& e^{-\mu} \nu,_{zz}+e^{-\mu} \nu,_z^2 +\frac 12 e^{-\mu} \nu,_r\mu,_r-\frac 12 e^{-\mu} \nu,_z\mu,_z -\frac 34 e^{2\psi-\mu-2\nu} \chi,_z^2,  \\
 \Omega_{0102}=& e^{-\mu} \nu,_{rz}+e^{-\mu} \nu,_z \nu,_r -\frac 12 e^{-\mu} \nu,_r\mu,_z -\frac 12 e^{-\mu} \nu,_z\mu,_r -\frac 34 e^{2\psi-\mu-2\nu} \chi,_z\chi,_r,   \\
 \Omega_{0113}=& -\frac 12 e^{-\frac \mu2} \partial_z(e^{\psi-\frac \mu2-\nu} \chi,_z)-\frac 14 e^{\psi-\mu-\nu} \chi,_r\mu,_r -e^{\psi-\mu-\nu} \chi,_z\psi,_z,   \\
 \Omega_{0123}=& -\frac 12 e^{-\frac \mu2} \partial_r(e^{\psi-\frac \mu2-\nu} \chi,_z)-\frac 12 e^{-\mu-\nu} \chi,_z\psi,_r+\frac 14 e^{\psi-\mu-\nu} \chi,_r\mu,_z -\frac 12 e^{\psi-\mu-\nu} \chi,_r\psi,_z,  \\
 \Omega_{0201}=& e^{-\mu} \nu,_{rz}+e^{-\mu} \nu,_r\nu,_z -\frac 12 e^{-\mu} \nu,_z\mu,_r -\frac 12 e^{-\mu} \nu,_r\mu,_z -\frac 34 e^{2\psi-\mu-2\nu} \chi,_z\chi,_r, \\
 \Omega_{0202}=& e^{-\mu} \nu,_{rr}+e^{-\mu} \nu,_r^2+\frac 12 e^{-\mu} \nu,_z\mu,_z-\frac 12 e^{-\mu} \nu,_r\mu,_r -\frac 34 e^{2\psi-\mu-2\nu} \chi,_r^2, \\
 \Omega_{0213}=& -\frac 12 e^{-\frac \mu2} \partial_z(e^{\psi-\frac \mu2-\nu} \chi,_r)-\frac 12 e^{\psi-\mu-\nu} \chi,_r\psi,_z+\frac 14 e^{\psi-\mu-\nu} \chi,_z\mu,_r -\frac 12 e^{\psi-\mu-\nu} \chi,_z\psi,_r,  \\
 \Omega_{0223}=& -\frac 12 e^{-\frac \mu2} \partial_r(e^{\psi-\frac \mu2-\nu} \chi,_r)-e^{\psi-\mu-\nu} \chi,_r\psi,_r-\frac 14 e^{\psi-\mu-\nu} \chi,_z\mu,_z,  \\
 \Omega_{0312}=& \frac 12 e^{-\frac \mu2} \partial_r(e^{\psi-\frac \mu2-\nu} \chi,_z)+\frac 14 e^{\psi-\mu-\nu} \chi,_z\mu,_r-\frac 12 e^{-\frac \mu2} \partial_z(e^{\psi-\frac\mu2-\nu} \chi,_r) -\frac 14 e^{\psi-\mu-\nu} \chi,_r\mu,_z,  \\
 \Omega_{0303}=& e^{-\mu}\nu,_z \psi,_z+\frac 14 e^{2\psi-\mu-2\nu} \chi,_z^2 +e^{-\mu} \nu,_r\psi,_r+\frac 14 e^{2\psi-\mu-2\nu}\chi,_r^2, \\
 \Omega_{1212}=& -\frac 12 e^{-\mu} \mu,_{rr}-\frac 12 e^{-\mu} \mu,_{zz} ,   \\
 \Omega_{1203}=& \frac 12 e^{\psi-\mu-\nu}\nu,_z \chi,_r-\frac 12 e^{\psi-\mu-\nu} \chi,_z\nu,_r +\frac 12 e^{\psi-\mu-\nu} \chi,_z\psi,_r-\frac 12 e^{\psi-\mu-\nu}\chi,_r\psi,_z, \\
 \Omega_{1301}=& -\frac 12 e^{-\frac \mu2} \partial_z(e^{\psi-\frac \mu2-\nu} \chi,_z)-e^{\psi-\mu-\nu} \chi,_z\psi,_z-\frac 14 e^{\psi-\mu-\nu} \chi,_r\mu,_r,  \\
 \Omega_{1302}=& -\frac 12 e^{-\frac \mu2} \partial_z(e^{\psi-\frac \mu2-\nu} \chi,_r)-\frac 12 e^{\psi-\mu-\nu} \chi,_r\psi,_z+\frac 14 e^{\psi-\mu-\nu} \chi,_z\mu,_r -\frac 12 e^{\psi-\mu-\nu} \chi,_z\psi,_r,  \\
 \Omega_{1313}=& -e^{-\frac \mu2} \partial_z(e^{-\frac \mu2} \psi,_z)- e^{-\mu} \psi,_z^2-\frac 14 e^{2\psi-\mu-2\nu} \chi,_z^2 -\frac 12 e^{-\mu} \mu,_r\psi,_r, \\
 \Omega_{1323}=& -e^{-\frac \mu2} \partial_r(e^{-\frac \mu2} \psi,_z)- e^{-\mu} \psi,_z\psi,_r-\frac 14 e^{2\psi-\mu-2\nu} \chi,_z\chi,_r +\frac 12 e^{-\mu} \mu,_z\psi,_r, \\
 \Omega_{2301}=& -\frac 12 e^{-\frac \mu2} \partial_r(e^{\psi-\frac \mu2-\nu} \chi,_z)-\frac 12 e^{\psi-\mu-\nu} \chi,_z\psi,_r+\frac 14 e^{\psi-\mu-\nu} \chi,_r\mu,_z -\frac 12 e^{\psi-\mu-\nu} \chi,_r\psi,_z, \\
 \Omega_{2302}=& -\frac 12 e^{-\frac \mu2} \partial_r(e^{\psi-\frac \mu2-\nu} \chi,_r)-e^{\psi-\mu-\nu} \chi,_r\psi,_r-\frac 14 e^{\psi-\mu-\nu} \chi,_z\mu,_z,  \\
 \Omega_{2313}=& -e^{-\frac \mu2} \partial_r(e^{-\frac \mu2} \psi,_z)- e^{-\mu} \psi,_z\psi,_r-\frac 14 e^{2\psi-\mu-2\nu} \chi,_z\chi,_r +\frac 12 e^{-\mu} \mu,_z\psi,_r, \\
 \Omega_{2323}=& -e^{-\frac \mu2} \partial_r(e^{-\frac \mu2} \psi,_r)- e^{-\mu} \psi,_r^2-\frac 14 e^{2\psi-\mu-2\nu} \chi,_r^2 -\frac 12 e^{-\mu} \mu,_z\psi,_z.
\end{align}
These simplify when restricting to the equatorial plane and assuming symmetry:
\begin{align}
 \Omega_{0101}=& \frac 12 e^{-\mu} \nu,_r\mu,_r,  \\
 \Omega_{0113}=& -\frac 14 e^{\psi-\mu-\nu} \chi,_r\mu,_r,   \\
 \Omega_{0202}=& e^{-\mu} \nu,_{rr}+e^{-\mu} \nu,_r^2-\frac 12 e^{-\mu} \nu,_r\mu,_r -\frac 34 e^{2\psi-\mu-2\nu} \chi,_r^2, \\
 \Omega_{0223}=& -\frac 12 e^{-\frac \mu2} \partial_r(e^{\psi-\frac \mu2-\nu} \chi,_r)-e^{\psi-\mu-\nu} \chi,_r\psi,_r,  \\
 \Omega_{0303}=& e^{-\mu} \nu,_r\psi,_r+\frac 14 e^{2\psi-\mu-2\nu}\chi,_r^2, \\
 \Omega_{1212}=& -\frac 12 e^{-\mu} \mu,_{rr},   \\
 \Omega_{1301}=& -\frac 14 e^{\psi-\mu-\nu} \chi,_r\mu,_r,  \\
 \Omega_{1313}=& -\frac 12 e^{-\mu} \mu,_r\psi,_r, \\
 \Omega_{2302}=& -\frac 12 e^{-\frac \mu2} \partial_r(e^{\psi-\frac \mu2-\nu} \chi,_r)-e^{\psi-\mu-\nu} \chi,_r\psi,_r,  \\
 \Omega_{2323}=& -e^{-\frac \mu2} \partial_r(e^{-\frac \mu2} \psi,_r)- e^{-\mu} \psi,_r^2-\frac 14 e^{2\psi-\mu-2\nu} \chi,_r^2. 
\end{align}
The Riemann tensor is then related to the coefficients of the curvature form by
\begin{align}
 R_{\mu\nu\rho\sigma}=e^a_\mu e^b_\nu e^c_\rho e^d_\sigma \Omega_{abcd}.
\end{align}
The independent, non-vanishing components of the Riemann tensor are
\begin{align}
    R_{tztz}=& \Omega_{0101}e^{2\nu+\mu}+2\Omega_{0113}e^{\nu+\mu+\psi}\chi+\Omega_{1313}e^{\mu+2\psi}\chi^{2}, \\
    R_{tztr}=& \Omega_{0102}e^{2\nu+\mu}+(\Omega_{0123}+\Omega_{0213})e^{\nu+\mu+\psi}\chi+\Omega_{1323}e^{\mu+2\psi}\chi^{2}, \\
    R_{tzz\phi}=& \Omega_{0113}e^{\nu+\mu+\psi}+\Omega_{1313}e^{\mu+2\psi}\chi, \\
    R_{tzr\phi}=& \Omega_{0123}e^{\nu+\mu+\psi}+\Omega_{1323}e^{\mu+2\psi}\chi, \\
    R_{trtr}=& \Omega_{0202}e^{2\nu+\mu}+2\Omega_{0223}e^{\nu+\mu+\psi}\chi+\Omega_{2323}e^{\mu+2\psi}\chi^{2}, \\
    R_{trz\phi}=& \Omega_{0213}e^{\nu+\mu+\psi}+\Omega_{2313}e^{\mu+2\psi}\chi, \\
    R_{trr\phi}=& \Omega_{0223}e^{\nu+\mu+\psi}+\Omega_{2323}e^{\mu+2\psi}\chi, \\
    R_{t\phi t\phi}=& \Omega_{0303}e^{2\nu+2\psi}, \\
    R_{t\phi zr}=& \Omega_{0312}e^{\nu+\mu+\psi}, \\
    R_{zrzr}=& \Omega_{1212}e^{2\mu}, \\
    R_{z\phi z\phi}=& \Omega_{1313}e^{\mu+2\psi}, \\
    R_{r\phi r\phi}=& \Omega_{2323}e^{\mu+2\psi}.
\end{align}
Restricting to the galactic plane and imposing symmetry with respect to it we obtain
\begin{align}
    R_{tztz}=& \Omega_{0101}e^{2\nu+\mu}+2\Omega_{0113}e^{\nu+\mu+\psi}\chi+\Omega_{1313}e^{\mu+2\psi}\chi^{2}, \\
    R_{tzz\phi}=& \Omega_{0113}e^{\nu+\mu+\psi}+\Omega_{1313}e^{\mu+2\psi}\chi, \\
    R_{trtr}=& \Omega_{0202}e^{2\nu+\mu}+2\Omega_{0223}e^{\nu+\mu+\psi}\chi+\Omega_{2323}e^{\mu+2\psi}\chi^{2}, \\
    R_{trr\phi}=& \Omega_{0223}e^{\nu+\mu+\psi}+\Omega_{2323}e^{\mu+2\psi}\chi, \\
    R_{t\phi t\phi}=& \Omega_{0303}e^{2\nu+2\psi}, \\
    R_{zrzr}=& \Omega_{1212}e^{2\mu}, \\
    R_{z\phi z\phi}=& \Omega_{1313}e^{\mu+2\psi}, \\
    R_{r\phi r\phi}=& \Omega_{2323}e^{\mu+2\psi}.
\end{align}

%%%%%%%%%%%%%%%%%%%%%%%%%%%%%%%%%%%%%%%

\section{Appendix C}

\label{A: Riemann locally inertial reference system}

The Riemann tensor in the local inertial frame is given by
\begin{equation}
    R_{abcd}=R_{\alpha\beta\mu\nu}e^{\alpha}_{a}e^{\beta}_{b}e^{\mu}_{c}e^{\nu}_{d},
\end{equation}
with $\{a,b,c,d\}\in\{0,1,2,3\}$ and $\{\alpha, \beta, \mu, \nu\}\in\{t, z, r, \phi\}$. The calculation is performed in the galactic plane and a prime denotes derivation with respect to $r$. There are twelve non-vanishing, independent components:
\begin{align}
    R_{0101} &= \frac{e^{-\mu} \mu' ( -e^{2 \nu} \nu' + e^{2 \psi} (\chi - \Omega) ( \chi' + (\chi - \Omega) \psi' ) )}{2 H}, \\ \notag \\
    R_{0112} &= \frac{1}{4rH}\left[e^{-\mu} \sin(\omega \tau) \mu' ( 2 e^{2 \nu} \eta \nu' - e^{2 \psi} ( (H + 2 \eta (\chi - \Omega)) \chi' \right.\notag \\
    &\left.+ 2 (H + \eta (\chi - \Omega)) (\chi - \Omega) \psi' ) )\right], \\ \notag \\
    R_{0113} &=\frac{1}{4rH} \left[e^{-\mu} \cos(\omega \tau) \mu' ( 2 e^{2 \nu} \eta \nu' - e^{2 \psi} ( (H + 2 \eta (\chi - \Omega)) \chi'\right. \notag \\
    &\left.+ 2 (H + \eta (\chi - \Omega)) (\chi - \Omega) \psi' ) )\right], \\ \notag \\
    R_{0202} &= \frac{1}{4r^{2}H}\left[e^{-\mu - 2 \nu} 
    (e^{4 \psi} r^2 \cos(\tau \omega)^2 (\chi - \Omega)^2 \chi'^{2} +\right. \notag\\ 
    &\left.+2 e^{2 \nu} r^2 \cos(\tau \omega)^2 \mu' (e^{2 \nu} \nu' - 
      e^{2 \psi} (\chi - \Omega) (\chi' + (\chi - \Omega) \psi')) + \right.\notag\\ 
    &\left.+4 e^{4 \nu} (e^{2 \psi} H \sin(\tau \omega)^2 \nu' \psi' - 
      r^2 \cos(\tau \omega)^2 (\nu'^{2} + \nu'' ) ) +\right. \notag\\ 
    &\left.+e^{2 (\nu + \psi)} (e^{2 \psi} H \sin(\tau \omega)^2 \chi'^{2} + 
      r^2 \cos(\tau \omega)^2 (3 \chi'^{2} + 4 \Omega (\chi' (\nu' - 3 \psi') - (\chi'') ) +\right. \notag\\ 
    &\left.+4 \chi^{2} (\psi'^{2} + (\psi'') ) + 4 \Omega^{2} (\psi'^{2} + (\psi'') ) + 4 
      \chi (-\nu' \chi' + 3 \chi' \psi' + (\chi'') \right.\notag\\
    &\left.- 2 \Omega (\psi'^{2} + (\psi'') )))))\right],
\end{align}
\begin{align}
    R_{0203} &=\frac{1}{8r^{2}H}\left[e^{-\mu - 2 \nu} \sin(2 \tau \omega) (-e^{4 \psi} r^{2} (\chi - \Omega)^{2} \chi'^{2}\right. + \notag\\ 
    &\left.+2 e^{4 \nu} (2 e^{2 \psi} H \nu' \psi' + r^{2} (-\mu' \nu' + 2 (\nu'^{2} + \nu'' )) +\right. \notag\\ 
    &\left.+e^{2 (\nu + \psi)} (e^{2 \psi} H \chi'^{2} + r^{2} (-3 \chi'^{2} - 2 \Omega \chi' (\mu' + 2 \nu' - 6 \psi') +\right.\notag \\ 
    &\left.+4 \Omega \chi'' + 2 \chi (2 \nu' \chi' - 6 \chi' \psi' + 4 \Omega \psi'^{2} + \mu' (\chi' - 2 \Omega \psi') - 2 \chi'' + 4 \Omega \psi'' ) \right.\notag\\
    &\left.+ 2 \chi^{2} (\mu' \psi' - 2 (\psi'^{2} + \psi'' )) +\right.\notag \\ 
    &\left.+2 \Omega^{2} (\mu' \psi' - 2 (\psi'^{2} + \psi'' ))))))\right], \\ \notag \\
    R_{0223} &= \frac{1}{4rH}\left[e^{-\mu + 2 \psi} \cos(\tau \omega) (e^{-2 \nu + 2 \psi} (H + \eta (\chi - \Omega)) (\chi - \Omega) \chi'^{2} +\right. \notag \\
    &\left.+2 e^{2 \nu - 2 \psi} \eta (\mu' \nu' - 2 (\nu'^{2} + \nu'')) + \eta (3 \chi'^{2} + 2 \Omega (\chi' (\mu' + 2 \nu' - 6 \psi') - 2 \chi'') - \right.\notag \\
    &\left.+2 \chi (2 \nu' \chi' - 6 \chi' \psi' + 4 \Omega \psi'^{2} + \mu' (\chi' - 2 \Omega \psi') - 2 \chi'' + 4 \Omega \psi'')) +\right. \notag \\
    &\left.+H (-\mu' (\chi' + 2 (\chi - \Omega) \psi') + 2 (-\nu' \chi' + 3 \chi' \psi' + \chi'' \right. \notag \\
    &\left.+ 2 (\chi - \Omega) (\psi'^{2} + \psi''))))\right], \\ \notag \\
    R_{0303} &= \frac{1}{4r^{2}H}\left[e^{-\mu - 2 \nu} ( e^{4 \psi} r^2 \sin^2(\tau \omega) (\chi - \Omega)^2 \chi'^{2}\right. + \notag \\
    &\left.+2 e^{2 \nu} r^2 \sin^2(\tau \omega) \mu' ( e^{2 \nu} \nu' - e^{2 \psi} (\chi - \Omega) ( \chi' + (\chi - \Omega) \psi' ) ) + \right.\notag \\
    &\left.+4 e^{4 \nu} ( e^{2 \psi} \cos^2(\tau \omega) H \nu' \psi' - r^2 \sin^2(\tau \omega) ( \nu'^{2} + \nu'' ) ) +\right. \notag \\
    &\left.+e^{2 (\nu + \psi)} ( e^{2 \psi} \cos^2(\tau \omega) H \chi'^{2} + r^2 \sin^2(\tau \omega) ( 3 \chi'^{2} + 4 \Omega ( \chi' (\nu' - 3 \psi') - \chi'' ) + \right.\notag \\
    &\left.+4 \chi^{2} ( \psi'^{2} + \psi'' ) + 4 \Omega^{2} ( \psi'^{2} + \psi'' ) + 4 \chi ( -\nu' \chi' + 3 \chi' \psi' + \chi'' \right. \notag \\
    &\left.- 2 \Omega ( \psi'^{2} + \psi'' ) ) ) ) )\right], \\ \notag \\
    R_{0323} &= \frac{1}{4rH}\left[e^{-\mu + 2 \psi} \sin(\tau \omega) ( e^{-2 \nu + 2 \psi} ( H + \eta (\chi - \Omega) ) (\chi - \Omega) \chi'^{2} +\right. \notag \\
    &\left.+2 e^{2 \nu - 2 \psi} \eta ( \mu' \nu' - 2 ( \nu'^{2} + \nu'' ) ) + \eta ( 3 \chi'^{2} + 2 \Omega ( \chi' ( \mu' + 2 \nu' - 6 \psi' ) - 2 \chi'' )+ \right.\notag \\
    &\left.-2 \chi ( 2 \nu' \chi' - 6 \chi' \psi' + 4 \Omega \psi'^{2} + \mu' ( \chi' - 2 \Omega \psi' ) - 2 \chi'' + 4 \Omega \psi'' ) + \right.\notag \\
    &\left.+\chi^{2} ( -2 \mu' \psi' + 4 ( \psi'^{2} + \psi'' ) ) + \Omega^{2} ( -2 \mu' \psi' + 4 ( \psi'^{2} + \psi'' ) ) ) \right. \notag \\
    &\left.+ H ( -\mu' ( \chi' + 2 (\chi - \Omega) \psi' ) +\right. \notag \\
    &\left.+2 ( -\nu' \chi' + 3 \chi' \psi' + \chi'' + 2 (\chi - \Omega) ( \psi'^{2} + \psi'' ) ) )\right], \\ \notag \\
    R_{1212} &=\frac{1}{2r^{2}H} \left[e^{-\mu} ( -e^{2 \nu} \sin(\tau \omega)^{2} \eta^{2} \mu' \nu' + \right.\notag \\
    &\left.+e^{2 \psi} \sin(\tau \omega)^{2} ( H + \eta (\chi - \Omega) ) \mu' ( \eta \chi' + ( H + \eta (\chi - \Omega) ) \psi' ) \right. \notag \\
    &\left.- r^{2} \cos(\tau \omega)^{2} H \mu'' )\right], \\ \notag \\
    R_{1213} &= \frac{1}{4r^{2}H}\left[e^{-\mu} \sin(2 \tau \omega) ( -e^{2 \nu} \eta^{2} \mu' \nu' + e^{2 \psi} ( H + \eta (\chi - \Omega) ) \right.\notag \\
    &\left.\times\mu' ( \eta \chi' + ( H + \eta (\chi - \Omega) ) \psi' ) + r^{2} H \mu'' )\right], 
\end{align}
\begin{align}
    R_{1313} &=\frac{1}{2r^{2}H}\left[ e^{-\mu} ( -e^{2 \nu} \cos^{2}(\tau \omega) \eta^{2} \mu' \nu' + +e^{2 \psi} \cos^{2}(\tau \omega) ( H + \eta (\chi - \Omega) ) \right.\notag \\
    &\left.\times\mu' ( \eta \chi' + ( H + \eta (\chi - \Omega) ) \psi' ) - r^{2} H \sin^{2}(\tau \omega) \mu'' )\right], \\ \notag \\ \\ \\
    R_{2323} &=\frac{1}{4r^{2}H}\left[ e^{-\mu - 2\nu} ( e^{4\psi} ( H + \eta (\chi - \Omega) )^{2} \chi'^{2} + 2 e^{4\nu} \eta^{2} ( \mu' \nu' - 2 ( \nu'^{2} + \nu'' ) )+\right. \notag \\
    &\left.- e^{2(\nu + \psi)} ( 2 H^{2} ( \mu' \psi' - 2 ( \psi'^{2} + \psi'' ) ) + \eta^{2} ( -3 \chi'^{2} - 2 \Omega \chi' ( \mu' + 2\nu' - 6\psi' ) + \right.\notag \\
    &\left.+4 \Omega \chi'' + 2 \chi ( 2 \nu' \chi' - 6 \chi' \psi' + 4 \Omega \psi'^{2} + \mu' ( \chi' - 2 \Omega \psi' ) - 2 \chi'' + 4 \Omega \psi'' ) + \right.\notag \\
    &\left.+2 \chi^{2} ( \mu' \psi' - 2 ( \psi'^{2} + \psi'' ) ) + 2 \Omega^{2} ( \mu' \psi' - 2 ( \psi'^{2} + \psi'' ) ) ) \right.\notag\\
    &\left.+ 2 H \eta ( \mu' ( \chi' + 2 (\chi - \Omega) \psi' )+\right. \notag \\
    &\left.- 2 ( - \nu' \chi' + 3 \chi' \psi' + \chi'' + 2 (\chi - \Omega) ( \psi'^{2} + \psi'' ) ) ) ) )\right].
\end{align}

For the BG model one finds
\begin{align}
    R_{0112}&= -\frac{e^{-\mu} \sin(\tau \omega) \eta' \mu'}{4r}, \\
    R_{0113}&=\frac{e^{-\mu} \cos(\tau \omega) \eta' \mu'}{4r}, \\
    R_{0202}&=\frac{e^{-\mu} \eta'^2}{4r^2}, \\
    R_{0223}&=-\frac{e^{-\mu} \cos(\tau \omega) \left( \eta' (2 + r \mu') - 2r \eta'' \right)}{4r^2}, \\
    R_{0303}&=\frac{e^{-\mu} \eta'^2}{4 r^2},\\
    R_{0323}&=-\frac{e^{-\mu} \sin(\tau \omega) \left( \eta' \left( 2 + r \mu' \right) - 2 r \eta'' \right)}{4 r^2}, \\
    R_{1212}&=-\frac{e^{-\mu} \left( \sin(\tau \omega)^2 \mu' + r \cos(\tau \omega)^2 \mu'' \right)}{2 r}, \\
    R_{1213}&=\frac{e^{-\mu} \sin(2 \tau \omega) \left( \mu' - r \mu'' \right)}{4 r}, \\
    R_{1313}&=-\frac{e^{-\mu} \left( \cos(\tau \omega)^2 \mu' + r \sin(\tau \omega r)^2 \mu''[r] \right)}{2 r}, \\
    R_{2323}&=\frac{e^{-\mu} \left( 3 \eta'^2 + 2 r \mu'6 \right)}{4 r^2}.
\end{align}

\clearpage

\bibliographystyle{unsrt}

\bibliography{bib.bib}

\end{document}